\documentclass[aip,reprint]{revtex4-1}

\usepackage{graphicx,nomencl}
\usepackage[fleqn]{amsmath}
\usepackage{bm}
\usepackage{float}
\usepackage{multirow}
\usepackage{array}

\usepackage{etoolbox}
\renewcommand\nomgroup[1]{%
  \item[\bfseries
  \ifstrequal{#1}{F}{Flow characteristics}{%
  \ifstrequal{#1}{C}{Cavitation modelling}{%
  \ifstrequal{#1}{D}{Dimensionless numbers}{}}}%
]}
\usepackage[xcdraw]{xcolor}
\newcommand{\red}[1]{#1}

\begin{document}

{
\scriptsize
\noindent

\onecolumngrid
This article may be downloaded for personal use only. Any other use requires prior permission of the author and AIP Publishing. This article appeared in \textit{Physics of Fluids 35, 081301 (2023)} and may be found at \url{https://doi.org/10.1063/5.0157926}.
}

\title{A classification and review of cavitation models with an emphasis on physical aspects of cavitation}

\author{Tobias Simonsen Folden}
\affiliation{Department of Mathematical Sciences, Aalborg University, A. C. Meyers Vænge 15, 2450 Copenhagen, Denmark }

\author{Fynn Jerome Aschmoneit}
\email[]{fynnja@math.aau.dk}
\affiliation{Department of Mathematical Sciences, Aalborg University, A. C. Meyers Vænge 15, 2450 Copenhagen, Denmark }

\date{\today}

\begin{abstract}
This review article presents a summary of the main categories of models developed for modelling cavitation, a multiphase phenomenon in which a fluid locally experiences phase change due to a drop in ambient pressure.
The most common approaches to modelling cavitation along with the most common modifications to said approaches due to other effects of cavitating flows are identified and categorized.
The application of said categorization is demonstrated through an analysis of selected cavitation models.
For each of the models presented, the various assumptions and simplifications made by the authors of the model are discussed, and applications of the model to simulating various aspects of cavitating flow are also presented.
The result of the analysis is demonstrated via a visualisation of the categorizations of the highlighted models.
Using the preceding discussion of the various cavitation models presented, the review concludes with an outlook towards future improvements in the modelling of cavitation.

\end{abstract}

\maketitle %\maketitle must follow title, authors, abstract and \pacs

% \nomenclature[F]{$\mathbf{u}$}{(local) velocity}
% % \nomunit{\qty{m^3/m^3}}
% \nomenclature[F]{$u$}{(local) speed}
% \nomenclature[F]{$u_\infty$}{characteristic speed}
% \nomenclature[F]{$p$}{(local) pressure}
% \nomenclature[F]{$p_v(T)$}{vapor pressure at temperature $T$}
% \nomenclature[F]{$\Delta p$}{characteristic pressure drop}
% \nomenclature[F]{$p_\infty$}{freestream pressure}
% \nomenclature[F]{$\rho$}{density}
% \nomenclature[F]{$\mu$}{viscosity}
% \nomenclature[F]{$k$}{turbulent kinetic energy}
% \nomenclature[F]{$\varepsilon$}{rate of dissipation of turbulent kinetic energy}
% \nomenclature[C]{$\sigma$}{surface tension}
% \nomenclature[C]{$R$}{bubble radius}
% \nomenclature[C]{$n_0$}{bubble number density per unit volume}
% \nomenclature[C]{$\alpha_i$}{volume fraction of the $i$th phase}
% \nomenclature[C]{$f_i$}{mass fraction of the $i$th phase}
% \nomenclature[C]{$\Dot{m}_i$}{mass transfer rate for the $i$th phase}
% \nomenclature[C]{$\Dot{m}^+$}{mass transfer rate during vaporisation}
% \nomenclature[C]{$\Dot{m}^-$}{mass transfer rate during condensation}
% \nomenclature[C]{$C_+$}{empirical constant used for modelling $\Dot{m}^+$}
% \nomenclature[C]{$C_-$}{empirical constant used for modelling $\Dot{m}^-$}
% \nomenclature[F]{$c$}{(local) speed of sound}
% \nomenclature[F]{$c_{wallis}$}{propagation of acoustic waves without mass transfer}
% \nomenclature[C]{$t_\infty$}{characteristic time scale}

% \nomenclature[D]{$Ca$}{Cavitation number}
% \nomenclature[D]{$Re$}{Reynolds number}
% \printnomenclature

\section{Introduction}{
    Cavitation is the phenomenon in which local regions of a fluid experience a phase transition from liquid to vapor as the ambient pressure drops below the fluid's vapor pressure\cite{francFundamentalsCavitation2010,brennenCavitationBubbleDynamics2014}.
    Cavitating flow describes a flow regime, where local pressure fluctuations cause the fluid to cavitate locally. 
    These cavities may form coherent bubbles attached to some surface (sheet cavities), transient eddies (vortex cavities), or dispersed bubbles (cloud cavities).
    Once a cavity is exposed to a higher pressure environment, it will change phase to the liquid state again.
    This phase change may be quite rapid, so that it is usually referred to as bubble implosion.
    These implosions create shock waves, which carry enough momentum to damage the confining material.
    
    Cavitating flows are found in various industrial applications, where they often pose problems for the respective technology.
    In hydrodynamic machines such as pumps, pressure exchangers or turbines, cloud cavitation may cause material erosion leading to machine failure, noise, vibrations or operation instabilities, such as head losses in pumps \cite{adamkowskiResonanceTorsionalVibrations2016, al-obaidiInvestigationEffectPump2019, wuUnsteadyFlowStructural2019, yeNumericalMethodologyCFD2021, sunNumericalInvestigationInterblade2020}.
    In ship propellers, sheet cavitation leads to erosion of the downstream side of the blades and it leads to reduced propulsion or propeller-hull vortex cavitation \cite{yilmazExperimentalNumericalInvestigation2020, wittekindPropellerCavitationNoise2016, petersNumericalPredictionCavitation2018, zhuNumericalInvestigationCavitation2012, melissarisCavitationErosionRisk2022}.
    In hydrofoils, sheet cavitation on the top side lead to a decrease in lift \cite{saitoNumericalAnalysisUnsteady2007, watanabeUnsteadyLiftDrag2014}.
    There are also industrial examples, where cavitating flows serve a distinct purpose.
    Acoustic cavitation is the principal mechanism behind sonochemistry, a method for surface cleaning. 
    The generation of imploding cavitation bubbles creates high frequency shock waves, which are used for surface cleaning. 
    This cleaning procedure is applied in various industrial applications, such as in ultrafiltration\cite{yusofPhysicalChemicalEffects2016}, in the food industry\cite{azamEfficacyUltrasoundTreatment2020}, or in industrial-scale heat exchangers \cite{kieserApplicationIndustrialScale2011}.
    It is therefore of great interest to understand and control cavitating flows, in order optimize the applications above to minimize or exploit cavitation efficiently.
    
    Cavitation is a microscopic effect, acting on much smaller temporal and spacial scales, compared to representative scales of the surrounding flow.
    From a macroscopic perspective, cavitation is affected by various variables: 
    Naturally, the static and dynamic pressure, as well as the local temperature govern the overall the cavitating flow.
    However, cavitating flow is tightly coupled to turbulent flow, as turbulent eddies cause local pressure fluctuations \cite{ohernExperimentalInvestigationTurbulent1990}, and also disperse coherent cavities into cloud cavities\cite{brandnerExperimentalInvestigationCloud2010, huangLargeEddySimulation2014}.
    As such, it is natural to consider the effects of turbulence when attempting to model cavitating flows; however, the task of implementing turbulent effects is complicated by the fact that the most widely used turbulence models were originally developed for single-phase flows, and the extension of turbulence models to multiphase flows is still an active area of research, with many distinct models such as the mixture $k-\varepsilon$ model by Behzadi et al.\cite{behzadiModellingDispersedBubble2004} proposed for modelling turbulence in multiphase flow.
    Furthermore, as cavities implode the resultant shock waves may cause secondary cavities to appear close by, which also implode, thus creating cascades of implosions \cite{vanrijsbergenHighSpeedVideoObservations2012, dularMechanismsCavitationErosion2015, melissarisCavitationErosionRisk2022}.
    Implosions in the direct vicinity of a surface don't create concentric shock waves, but rather produce a directed pressure pulse towards the surface, posing the primary cause for material damage \cite{mihatschCavitationErosionPrediction2015}.
    The implosion intensity and the bubble interactions in the cavity cloud are dependant on the phases' viscosities and their surface tension.
    Hence, cavitating flow entails the highly complex interactions between microscopic cavitation and the macroscopic flow.

    Due to the wide range of examples of cavitating flows with highly differing characteristics such as the geometry of the domain, the thermodynamic variables of the cavitating fluid, and the structure of the flow, there is currently neither a universal model for cavitation nor a general framework for approaching the problem of developing a model.
    Since cavitation was first recognized as a distinct phenomenon in fluid dynamics, various authors have derived models that seek to explain the mechanisms behind the phase transfer that occurs in cavitating flows by directly attempting to simulate the mass transfer rates between the liquid phase and the vapor phase, simulating the formation, growth, motion, and collapse of cavities within the flow, or a combination of these approaches.
    Through an investigation of the various techniques employed by the authors of the models, the various cavitation models can broadly be categorized according to the approaches used for simulating cavitation in the given model.
     
    Several authors have previously reviewed various aspects of cavitation modelling.
    Utturkar et al.\cite{utturkarRecentProgressModeling2005}, Luo et al.\cite{luoReviewCavitationHydraulic2016}, and Li et al.\cite{liCavitationModelsThermodynamic2021} analysed cavitation models for the specific applications of rocket propulsion, hydraulic machinery, and organic Rankine cycles, respectively. Niedzwiedzka et al.\cite{niedzwiedzkaReviewNumericalModels2016} wrote a review on homogeneous cavitation models, comparing how fundamental empirical parameters governing the phase change are expressed in various articles. 
    Models for bubble implosions and their erosive potential on surfaces are reviewed in Wang et al.\cite{wangDynamicsCavitationStructure2017}.

    This review of cavitation modelling supplements the previous reviews mentioned above, highlighting the major categories of cavitation models and the methodologies and physical models used to develop them, with the discussion of the models centered on the governing equations as well as the expressions constructed for various source terms used in the model.
    As a product of this analysis and discussion of cavitation modelling, a categorization of the various approaches to cavitation modelling is proposed, with models based on the same approach grouped into one category of models.
    \red{This categorization is developed with the intent of providing a tool capable of identifying an appropriate model for the given effects of cavitation an engineer or a researcher may wish to account for in their studies, while at the same time illustrating the complexity of said model by identifying how many distinct modelling approaches are employed in this model.
    }

    \red{The structure of this review is given as follows:
    Section \ref{sec:categorization} introduces the model categorization, with each level of the categorization presented in its own subsection.
    Section \ref{sec:analysis} presents an application of said categorization to a selection of cavitation models proposed by various of authors, using an analysis of said models as justification for the proposed categorization.
    Section \ref{sec:conclusion} presents a visual representation of the categorized models in the form of Venn diagrams as well as observations regarding the state of cavitation modelling based on the analysis performed in Section \ref{sec:analysis}.
    Lastly, Section \ref{sec:outlook} provides an outlook towards future developments in cavitation modelling on the basis of both the proposed categorization as well as other recent research directions within cavitation modelling.}

    \section{Categorization of Cavitation Models}\label{sec:categorization}
    As discussed in previous reviews of the literature, e.g. Li et al. \cite{liCavitationModelsThermodynamic2021}, the problem of modelling cavitation has been has been treated using a variety of approaches to simulating cavitation.
    Due to cavitation being a phenomenon in multiphase flows, a model for simulating cavitation must provide both an approximation of the flow governed by the Navier-Stokes equations
    \begin{equation}
        \begin{split}
            \frac{\partial \rho \mathbf{u}}{\partial t} +  \nabla\cdot (\rho \mathbf{u}\otimes \mathbf{u}) &= \nabla \cdot( \mu (\nabla \mathbf{u} + (\nabla \mathbf{u})^T ) )-\nabla p,\\
            \frac{\partial \rho}{\partial t} + \nabla \cdot (\rho \mathbf{u} ) &= 0,
        \end{split}\label{eq:NSE}
    \end{equation}
    but also an appropriate scheme for estimating the formation, growth, motion, shape, and collapse of the cavities present in the flow using the flow characteristics.
    In order to obtain more accurate simulations that better reflect the behavior of real cavitating flows, the model should be able to account for effects known to influence the growth rate of cavities or events involving mutual interaction between distinct cavities.
    In the remainder of this section, we introduce the most common approaches and effects employed in cavitation modelling, 
    \subsection{Categorization by modelling approach}\label{subsec:approach}{
    A common method for modelling the distinct phases in cavitating flows is to assume that the liquid phase and the vapor phase are in  mechanical and thermal equilibrium with the same velocity and pressure fields, and that the fluid characteristics such as the density and viscosity are assumed to be specified locally as a homogeneous mixture of the corresponding characteristics of the two pure phases with the mixing ratio $\alpha_v$, i.e.
    \begin{equation}
    \begin{split}
    \label{eq:mixtureRhoMu}
        \rho=\rho_m &= \alpha_v \rho_v + (1-\alpha_v) \rho_l, \\
        \mu=\mu_m  &= \alpha_v \mu_v + (1-\alpha_v) \mu_l,
    \end{split}
    \end{equation}
    where $\alpha_v$ is the volume fraction field of the vaporous phase, i.e. the ratio of the fluid volume occupied by vapor and the total fluid volume.
    The vapor volume fraction $\alpha_v$ can be used to give an approximation of distribution of cavities within the flow by discretizing the computational domain into a number of computational cells, then approximating $\alpha_v$ locally within each cell.
    Models based on this approach are known as homogeneous mixture models, and they combine a scheme for the Navier-Stokes equations \eqref{eq:NSE} with a scheme that incorporates the hypothesis \eqref{eq:mixtureRhoMu} in some way.
    Fig. \ref{fig:vof} illustrates the limits of the homogeneous mixture approach: it is tightly coupled with the discretization of the computational domain. 
    The top row of figures indicates a cavitating flow, in which a large cavity is dispersed under the action of turbulence, creating a cavity cloud.
    In the middle row, the same flow field is illustrated with the vapor volume fraction $\alpha_v$. 
    It is seen that the cavity is reasonably well resolved in the left figure, but as the cavity is dispersed, the vapor volume fraction cannot distinguish between individual bubbles any more.
    The cavity cloud on the right is only represented as a near-homogeneous field, cavities become absolutely indistinguishable.
    In the bottom row, one possible model for the bubble density is illustrated, where the numbers per control volume increase, as the the vapor volume fraction cannot capture the bubbles any more. 
    It therefore acts as a support field, when the computational grid fails to resolve bubbles.
    This macroscopic view on cavitating flow highlights the importance of well-developed cavitation models on the sub-grid scale.
    
    \begin{figure}[h]
    \centering
        \includegraphics[width=\linewidth]{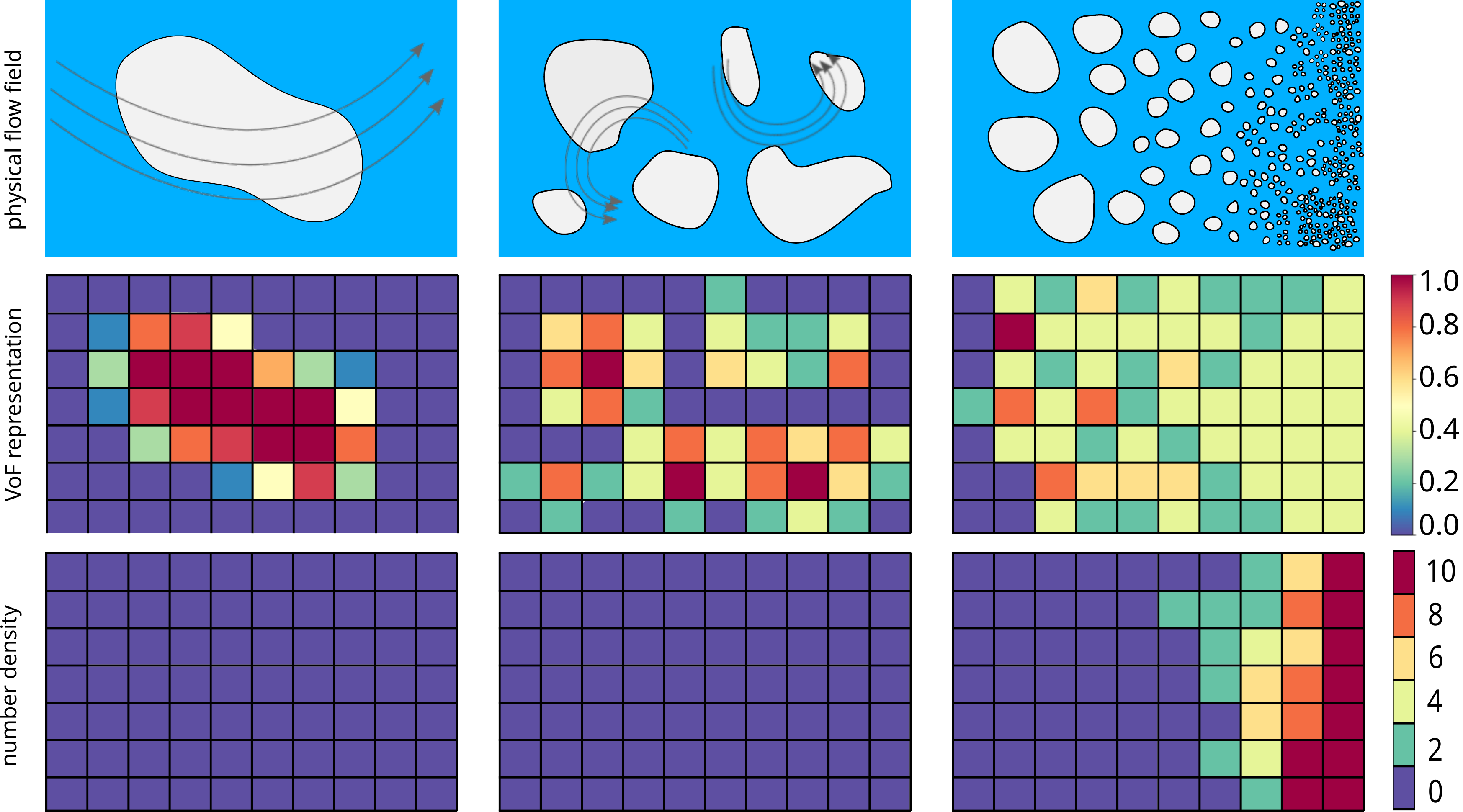}%
        \caption{\label{fig:vof} Cavitating flows undergoing turbulent bubble dispersion. (top) Illustration of a continuous flow, in which a large cavity is broken up into smaller cavities, forming a cavitation could. (middle) Volume-of-Fluid representation of the the bubble break up. The colors indicate the value of the vapor volume fraction. (bottom) A possible number density field, supplementing the VoF method as the computational grid resolution does not resolve bubbles any more. }%
    \end{figure}
    One approach to modelling cavitation using the homogeneous mixture approach as described by \eqref{eq:mixtureRhoMu} is to use a modified volume-of-fluid approach, which models the evolution of the vapor volume fraction $\alpha_v$ and the liquid volume fraction $\alpha_l=1-\alpha_v$ as described by the transport equations with source terms
    \begin{equation}
    \begin{split}
    \label{eq:vof}
        &\frac{\partial \rho_i \alpha_i}{\partial t} + \nabla\cdot( \rho_i \alpha_i\mathbf{u}) = \Dot{m}_i.
    \end{split}
    \end{equation}
    The models based on this approach are referred to as transport equation models, or TEMs for short.
    With no further assumptions other than the homogeneous mixture assumption, it is sufficient to solve \eqref{eq:vof} for the volume fraction of only one phase, typically the vapor phase, as the liquid volume fraction trivially follows from the volume conservation.
    Based on these observations, we have adopted the convention of always considering the vapor phase whenever we discuss volume fractions and mass transfer rates throughout this review.
    The volume fraction source term $\Dot{m}_i$ in \eqref{eq:vof}, hereafter denoted by $\Dot{m}$, plays a crucial role in TEMs for cavitation and is the main distinguishing feature of said models, as this term gives an explicit rate of the phase transition.
    In order to account for the different physical aspects governing growth and collapse of cavitation bubbles, $\Dot{m}$ is commonly split into two distinct terms $\Dot{m}^+$ and $\Dot{m}^-$ called the evaporation term and the condensation term, respectively.
    The split of the source term is defined such that at any time, only one of $\Dot{m}^+$ and $\Dot{m}^-$ is active in each control volume based on the relationship between the local pressure $p$ and the vapor pressure $p_v$.
    When the pressure drops below the vapor pressure, the evaporation term $\Dot{m}^+$ is active; correspondingly, the condensation term $\Dot{m}^-$ is active whenever the pressure exceeds the vapor pressure.
    Mathematically, the split of the source term can be expressed as
    \begin{align}
        \Dot{m}=\begin{cases}
            \Dot{m}^+, & p<p_v,\\
            \Dot{m}^-, & p>p_v.
        \end{cases}\label{eq:source_term_split}
    \end{align}
    The main challenge in using the TEM approach is developing an appropriate expression for the source terms $\Dot{m}^+$ and $\Dot{m}^-$, which is done through an additional hypothesis relating the vapor volume fraction to some other quantity whose growth can be estimated.
    The most commonly used approach to develop an expression for the source term is to assume that within each computational cell of the domain, the vapor volume fraction is approximately equal to the amount of volume of the cell occupied by bubbles of vapor that form from microscopic vapor nuclei which grow and collapse as the local pressure experienced by these bubbles fall below and exceed the vapor pressure $p_v(T)$.
    The growth and collapse of these bubbles is modelled by expressing their radial growth via the Rayleigh-Plesset equation\cite{rayleighPressureDevelopedLiquid1917,plessetDynamicsCavitationBubbles1949}, referred to as the RPE.
    This equation expresses the growth rate of a bubble with radius $R$ immersed in an infinite domain of liquid as
    \begin{equation}
    \label{eq:rayleighPlesset}
        R \frac{d^2 R}{dt^2} + \frac{3}{2} \left( \frac{dR}{dt}  \right)^2 = \frac{p_b(t) -p_\infty(t) }{\rho_l} -\frac{4 \mu_l}{\rho_l R} \frac{dR}{dt} - \frac{2 \sigma}{\rho_l R},
    \end{equation}
    which describes the evolution of the radius $R$ of a single spherical bubble immersed in an infinite expanse of liquid with internal pressure $p_b$ under the influence of a reference pressure $p_\infty$, the liquid viscosity $\mu_l$, and the liquid-bubble interface's surface tension $\sigma$.
    The full RPE is rarely used in cavitation modelling; instead, a simplified version obtained by neglecting e.g. the surface tension or the inertial effects is used in order to simplify calculations.
    Assuming that the cavity within a given control volume is approximately a cluster of spherical bubbles of the same size with radius $R$, the vapor volume fraction can be approximated via the bubble radius $R$ as
    \begin{equation}
        \alpha_v = n_0 \frac{4}{3} \pi R^3;\label{eq:cavmodel_vaporvolfrac_expr}
    \end{equation}
    in \eqref{eq:cavmodel_vaporvolfrac_expr}, $n_0$ is the bubble number density, which is either a constant parameter throughout the entire domain or another flow variable governed by its own transport equation depending on the model.
    The usage of the RPE as an approach to modelling cavitation is due to the work of Kubota et al.\cite{kubotaNewModellingCavitating1992}, who introduced these ideas in their cavitation model.
    Notably, Kubota et al. tracked the motion of a cluster of bubbles entirely without solving a transport equation for the volume fraction.

    Another approach to modelling cavitation within the homogeneous mixture framework is to develop an appropriate equation of state (EOS) that defines the thermodynamic behavior of cavitating flows and provides an expression that can be used to simulate the evolution of a quantity that is representative of the distribution of the two phases, most frequently the density $\rho$.
    The exact EOS used in these models highly depends on the nature of the fluid in question as well as the properties of the surrounding domain; models for hydrodynamic cavitation in the literature commonly employ a purely polytropic EOS, whilst models for acoustic cavitation make use of a more general EOS such as the stiffened gas EOS employed by Denner et al.\cite{dennerModelingAcousticCavitation2020}
    Additionally, there have also been made efforts to include an EOS in cavitation models primarily rooted in the TEM approach, e.g. the "four-equation" model by Goncalv\`es and Charri\`ere \cite{goncalvesModellingIsothermalCavitation2014}.
    The main advantage of employing an EOS is sidestepping the numerous hypotheses associated with the RPE, e.g. cavities being approximately spherical in shape and immersed in an infinite domain of liquid.
    Another benefit of introducing an EOS is a very direct inclusion of thermodynamic effects, enabling the development of cavitation models for cryogenic fluids.
    
    TEMs have been applied to a wide variety of problems concerned with simulating cavitating flows in the literature.
    For the problem of simulating larger cavity structures such as sheet cavities over e.g. hydrofoil, various authors have reported results that agree with previous experimental data, see e.g. \cite{gnanaskandanLargeEddySimulation2016, budichNumericalSimulationAnalysis2018}.
    However, other works have indicated that models such as TEMs, with their simplified models to frequently assume no interactions between bubbles, can fail to resolve the subgrid cavities when applied to the problem of simulating cloud cavities\cite{asnaghiNumericalExperimentalAnalysis2018} or greatly overestimate the bubble growth rate when the bubble-bubble interactions are accounted for\cite{yeModelingHydrodynamicCavitating2016}.
    This has lead to an interest in alternative approaches to modelling cavitation that remedies these flaws.
    
    One approach has been to attempt to track the interface(s) between the liquid phase and the vapor phase by locating contours of the form $p_v=c$ for some prescribed constant $c$, leading to the category of \textit{interface tracking models}.
    These methods models are named after their approach to modelling cavitation, which involves estimating the shape of the cavity interface and its evolution based on the behaviour of flow characteristics near the interface, e.g. estimating the shape of the cavity by locating regions in which the local temperature experiences a sharp decrease as in the model of Deshpande et al.\cite{deshpandeNumericalModelingThermodynamic1997}.
    This approach for modelling cavitation works well when modelling sheet cavitation in two-dimensional flows over e.g. hydrofoils, but faces difficulties associated with properly modelling the cloud cavitation occurring in the wake region of the cavity.
    Additionally, the interface tracking methods have faced difficulties in adapting their approach to modelling three-dimensional cavitating flows, leading to these methods being less developed than the previous two methods\cite{liCavitationModelsThermodynamic2021}.
    
    A more elaborate approach for resolving microscale cavitation effects within the homogeneous mixture framework is given by the \textit{Lagrangian models}, where the liquid phase is treated in the Eulerian framework of conservation equations, whilst the vapor phase is instead modelled by describing the Newtonian motion of individual bubbles or parcels of bubbles within the Lagrangian framework.
    These models have the capability of implementing the effects that are neglected in models such as TEMs, e.g the influence of non-condensable gas, the effect of turbulence at the sub-grid scales, and bubble-bubble interactions.
    As a trade-off, the requirement that the dynamics of each individual bubble be tracked limits the applicability of Lagrangian models for large-scale industrial problems; some authors have reported success in their efforts to apply Lagrangian models to such problems\cite{giannadakisModellingCavitationDiesel2008}.
    For small-scale problems, the Lagrangian models have seen more widespread usage, with several authors reporting results that agree with known experimental data\cite{fusterModellingBubbleClusters2011,maedaModelingExperimentalAnalysis2015}.
    
    Observing the difficulties of properly modelling cavitation at very small scales using models framed in the Eulerian framework as well as the difficulty of modelling large-scale cavitating flows with Lagrangian models, several authors have sought to combine the two approaches into one hybrid model, leading to the category of \textit{multiscale models}.
    Multiscale models are cavitation models explicitly developed to be capable of simulating the growth and collapse of both cavities of a size sufficiently large to be visibly detected, but also the effects of cavities existing on a scale too small to be properly resolved by conventional methods employed in models such as e.g. TEMs.
    These models track all cavities in the flow and classify them according to the local grid's capability to properly resolve their dynamics as either macro- or micro-scale cavities.
    A multiscale model can essentially be viewed as a combination of three models:
    \begin{enumerate}
        \item a model for the growth of the macro-scale cavities, usually a TEM,
        \item a model for the growth and motion of the micro-scale bubbles, tracking each the growth and motion of each bubble separately,
        \item a scheme for determining whether a macro-scale cavity/micro-scale bubble should transfer from being treated in its current macro-/micro-scale model to the other model, based on the computational grid's capacity to properly resolve the cavity/bubble in question.
    \end{enumerate}
    These models track all cavities within the flow, using two separate schemes for modelling cavitation depending of the size of the cavity: for larger singular cavities or clouds of bubbles, a model based on the homogeneous mixture approach is employed, with most multiscale models favoring a TEM for this purpose.
    At the same time, the motion and size of bubbles on a given characteristic length scale formed from nuclei present in the flow or breakup of larger bubbles is tracked using governing equations formulated in a Lagrangian framework; this enables simulation of cavitation at the smallest length scales.
    The two scales are related to each other through an appropriate scheme for determining when a given cavity is not resolved properly by the current scale and transitioning said cavity from its current scale to the other.
    In recent years, different multiscale models have been proposed by various authors, e.g. the models by Hsiao et al.\cite{hsiaoMultiscaleTwophaseFlow2017} and Ghahramani et al. \cite{ghahramaniComparativeStudyNumerical2019}, both of which apply distinct approaches in constructing their models.
    Furthermore, there has also been extensive effort devoted to studying the various aspects of multiscale modelling such as the influence of various model parameters and choices of discretizations on the performance of the model as well as the capability of the model to properly resolve the micro-scale dynamics\cite{liVeryLargeEddy2021,liMultiscaleModelingTipleakage2021,liLargeEddySimulation2021,wangEulerLagrangeStudy2021,wangNumericalPredictionCavitation2022}.
    Multiscale models show promise in regards to the problem of constructing a cavitation model that more accurately simulates the behaviour of cavitating flows, as seen from the results of Wang et al. \cite{wangInvestigationMultiscaleFeatures2023}, who performed a comparative study in order to investigate the characteristics of both a TEM and a multiscale model when applied to the case of simulating cavitating flow in a funnel over a raised flat section.
    Wang et al. found that the multiscale model exhibits both better agreement with experimental data for the validation case compared to the TEM and less sensitivity to the resolution of the computational mesh.
    
    On the basis of the modelling approaches introduced above, a cavitation model may be categorized according to the approaches employed within the model is introduced.
    The model is categorized as belonging to a combination of the following five categories:
    \begin{itemize}
        \item Category Rayleigh-Plesset Equation (RPE), where the model employs bubble dynamics as expressed via a (simplified) RPE \eqref{eq:rayleighPlesset},
        \item Category Transport Equation Model (TEM), where the model employs a transport equation to approximate the mass transfer rates and thus the growth and collapse of cavities,
        \item Category Equation Of State (EOS), where the model employs an EOS to relate the vapor volume fraction to thermodynamic variables,
        \item Category Interface Tracking Model (ITM), where the model employs a scheme that directly tracks the interface between the liquid and the vapor phase,
        \item Category Multiscale (MUL), where the model employs several schemes for simulating the growth and collapse of cavities at various length scales.
    \end{itemize}
    \subsection{Categorization by model effects}\label{subsec:effects}
    In addition to the categorization of models according to the employed approach described in section \ref{subsec:approach}, another point of interest regarding cavitation modelling is the additional effects of cavitation most frequently accounted for in the literature.
    These include adjustments to the expressions used for simulating e.g. mass transfer rates and hypotheses regarding the influence of phenomena such as turbulence on the cavitation process.
    
    Almost all models include additional parameters in their expressions used for e.g. mass transfer rates whose values are not based on any physical considerations, but instead adjusted empirically in order to provide a better fit between the model's predictions of cavitation and experimental data obtained from observations of a specified system.
    These parameters should ideally be chosen to provide the greatest fit under the operating conditions of the system at hand, e.g. shape of the domain and flow rate; however, the empirical nature of these constants makes the possibility of applying a set of empirical parameters that provide a good fit for one system to modelling another system questionable, and the problem of calibrating these empirical parameters for optimal performance is still an open problem.
    Some authors have proposed various approaches towards calibrating empirical constants, e.g. the random forest-based workflow proposed by  Sikirica et al. \cite{sikiricaCavitationModelCalibration2020}, but as with other aspects of cavitation modelling, no uniform approach exists as of yet.

    As discussed by O'Hern\cite{ohernExperimentalInvestigationTurbulent1990}, Brandner et al.\cite{brandnerExperimentalInvestigationCloud2010}, and Huang and Wang \cite{huangLargeEddySimulation2014} among others, the presence of turbulence in the system affects the cavitation process through both local fluctuations of pressure yielding pressure drops of sufficiently high scales to allow the inception of cavitation and through formed cavities being bombarded with the eddies carried by the turbulent flow, causing deformations in the surface of the cavity which may lead to the cavity rupturing and breaking up into a cloud of smaller cavities.
    As such, a proper turbulence model capable of accounting for the multiphase nature of cavitating flow is a necessity.

    In relation to the effects caused by turbulence, the population balance of cavities in the flow undergoes rapid changes due to both existing cavities breaking up into cavities of smaller sizes, cavities colliding and remaining in contact for long enough in order for coalescence to occur, cavities being formed at impurities on the surface of the domain and then entrained into the flow, and cavities entering or exiting the flow at the inlet and outlet of the system, respectively.
    In order to track these effects, cavitation models can be extended with either a model for solving the population balance equation associated with the bubble number density or devise schemes for detecting the occurrence of events that may lead to changes in the population balance based on specified criteria, then resolving these events and their impact on the distribution of cavities.

    Another possible effect to include is the possible presence of a third phase in the cavitating flow, in the form of non-condensable gas.
    Should this third phase be present, it has an effect on the approach used for modelling cavitation.
    For example, the homogeneous mixture hypothesis \eqref{eq:mixtureRhoMu} should be reformulated to include the amount of volume occupied by the non-condensable gas as expressed via its volume fraction $\alpha$, i.e.
    \begin{equation}
    \begin{split}
    \label{eq:mixtureRhoMu_gas}
        \rho=\rho_m &= \alpha_v \rho_v + \alpha_{ng}\rho_{ng}  + (1-\alpha_v-\alpha_{ng}) \rho_l, \\
        \mu=\mu_m  &= \alpha_v \mu_v + \alpha_{ng}\mu_{ng} + (1-\alpha_v-\alpha_{ng}) \mu_l.
    \end{split}
    \end{equation}
    Similarly, the internal bubble pressure $p_b$ used in the RPE \eqref{eq:rayleighPlesset} also depends on the pressure of the gas inside the bubble.

    Based on these considerations, we extend the categorization presented in section \ref{subsec:approach} with a second categorization according to the effects accounted for in the construction of the model.
    This categorization consists of four categories, defined as follows:
    \begin{itemize}
        \item Category Turbulence (TUR), where the model accounts for the effects of turbulence in their approach to modelling cavitation,
        \item Category Population Balance (POP), where the model accounts for changes in the cavity population as described by the bubble number distribution,
        \item Category Empirical (EMP), where the model includes empirical constants adjusted to the characteristics of a given flow,
        \item Category Non-Condensable Gas (NCG), where the model accounts for the presence of a third phase in the form of a non-condensable gas in the flow.
    \end{itemize}
    
}

\section{Classification of Selected Models}\label{sec:analysis}

Having categorized the most common approaches to cavitation modelling as well as the effects most frequently taken into consideration in the construction of a model for cavitation, this section is dedicated to an application of said categorization.
A collection of 20 cavitation models from across the literature are highlighted in this section for analysis and classification.
The highlighted models include the following:
\begin{enumerate}
    \item The Bubble Cluster Model by Kubota et al.\cite{kubotaNewModellingCavitating1992}
    \item The Thermodynamic Variable Table Model by Ventikos and Tzabiras\cite{ventikosNumericalMethodSimulation2000}
    \item The Lattice-Boltzmann EOS Model by Banerjee and Saritha\cite{banerjeeNumericalStudyCavitation2015}
    \item The Polytropic Closure Model by Denner et al. \cite{dennerModelingAcousticCavitation2020}
    \item The Ginzburg-Landau Potential Model by Kunz et al.\cite{kunzPreconditionedNavierStokes2000}
    \item The Interface Mass and Normal Momentum Model by Senocak and Shyy\cite{senocakInterfacialDynamicsbasedModelling2004}
    \item The Thermodynamic Interface Model by Deshpande et al.\cite{deshpandeNumericalModelingThermodynamic1997}
    \item The Wake Closure Model by Liu et al. \cite{liuNumericalMethodSimulation2006}
    \item The Bubble Density-Liquid Volume Coupling Model by Schnerr and Sauer\cite{schnerrPhysicalNumericalModeling2001}
    \item The Full Cavitation Model by Singhal et al.\cite{singhalMathematicalBasisValidation2002}
    \item The Vapor Nuclei-Adjusted Model by Zwart et al.\cite{zwartTwophaseFlowModel2004}
    \item The Viscosity-Oriented Model by Konstantinov et al.\cite{konstantinovNumericalCavitationModel2014}
    \item The Plane Surface Evaporation Model by Saito et al.\cite{saitoNumericalAnalysisUnsteady2003}
    \item The Four-Equation Model by Goncalv\`es and Charri\`ere\cite{goncalvesModellingIsothermalCavitation2014}
    \item The Ghost-Fluid Multiscale Model by Hisao et al.\cite{hsiaoMultiscaleTwophaseFlow2017}
    \item The Density-Based Convex Combination Model by Huang and Wang \cite{huangModifiedDensityBased2011}
    \item The Heat Balance Model by Shi et al.\cite{shiRayleighPlessetBasedTransport2014}
    \item The Population Balance Model by Li and Carrica\cite{liPopulationBalanceCavitation2021}
    \item The Euler-Lagrangian Multiscale Model by Ghahramani et al. \cite{ghahramaniNumericalSimulationAnalysis2021}
    \item The Stochastic Field Model by Dumond et al. \cite{dumondStochasticfieldCavitationModel2013}
\end{enumerate}

\red{
All models discussed this section are validated against experimental data of some sort.
Since different models may be validated against the same experimental data, a list of all experimental studies used for validation is found in Table \ref{tab:expCases}.
The models are presented in order of increasing complexity of their modelling approach in the sense that models belonging to only a single category within the categorization presented in section \ref{subsec:approach} are presented first, followed by models belonging to two categories, and so forth.
As a conclusion to the analysis of each highlighted model, a summary of recent studies applying the model and their findings are also presented.
For completeness, it is also highlighted under which discretization schemes the model were developed.
Note that physical models are ideally not limited to specific discretization schemes.
Some models have become standard methods in commercial or open-source CFD packages.
Following the analysis of all 20 models, their placement within the various categories considered in sections \ref{subsec:approach} and \ref{subsec:effects} are demonstrated visually using Venn diagrams in Figures \ref{fig:cav_model_venn} and \ref{fig:cav_modelEffects_venn}.
}

\begin{table}
    \centering
    \begin{tabular}{lll}
    \hline\hline
     \textbf{id}  &  \textbf{authors}  & \textbf{comment}   \\
    \hline
    \multicolumn{3}{l}{\textbf{Hydrofoil}}\\
     e1 \quad\quad      &   Izumida et al.\cite{izumidaRelationshipCharacteristicsPartial1980}    &           -          \\
     e2       &   Kubota et al.\cite{kubotaUnsteadyStructureMeasurement1989}
                                                                                          &         -          \\
    e3       &   Avellan et al.\cite{avellanGenerationMechanismDynamics1988}            &           -          \\
    e7       &   Shen \& Dimotakis\cite{shenInfluenceSurfaceCavitation1989}             &           -          \\
    e8       &   Hord et al.\cite{hordCavitationLiquidCryogens1973}                     &           Airfoil           \\
    e9       &   Keller \& Arndt\cite{kellerCavitationScaleEffects2001}                 &           -         \\
    e14       &   Berntsen et al.\cite{berntsenNumericalModelingSheet2001}              &           -         \\
    e15       &   Wang et al.\cite{wangDynamicsAttachedTurbulent2001}                   &           -         \\
    e17       &   Foeth\cite{foethStructureThreedimensionalSheet2008}                   &           -           \\
    \multicolumn{3}{l}{\textbf{Venturi nozzle}}\\
     e6       &   Stutz \& Reboud\cite{stutzTwophaseFlowStructure1997, stutzExperimentsUnsteadyCavitation1997,         
                                        stutzMeasurementsUnsteadyCavitation2000}        &           -           \\
    e12       &   Barre et al.\cite{barreExperimentsModelingCavitating2009}             &           -          \\
    e13       &   Patella et al.\cite{patellaExperimentsModellingCavitating2006}       &           -           \\
    \multicolumn{3}{l}{\textbf{Axis-symmetric}}\\
    e5       &   Rouse \& McNown\cite{rouseCavitationPressureDistribution1948}          &          Solid heads,  various forms \\    
    e16       &   Sarósdy \& Acosta\cite{sarosdyNoteObservationsCavitation1961}       &           Axisymmetric ogive           \\
    e18       &   Ghahramani et al.\cite{ghahramaniExperimentalNumericalStudy2020}       &         Semi-circular cylinder           \\
    \multicolumn{3}{l}{\textbf{other}}\\
    e4       &   Reuter \& Kaiser\cite{reuterHighspeedFilmthicknessMeasurements2019}    &           Bubble near solid wall           \\
    e10       &   Bakir et al.\cite{bakirNumericalExperimentalInvestigations2004}       &           Inducer           \\
    e11       &   Acekeret\cite{ackeretExperimentelleUndTheoretische1930}               &           Jet element           \\
    e19       &   Winklhofer et al.\cite{winklhoferComprehensiveHydraulicFlow2001}       &           Throttle           \\
     \hline\hline
    \end{tabular}
    \caption{Table of experimental cases used for model validation.}
    \label{tab:expCases}
\end{table}

\subsection{The Bubble Cluster Model}
    The use of bubble dynamics for modelling cavitation can be traced back to the Bubble Cluster Cavitation Model  by Kubota et al.\cite{kubotaNewModellingCavitating1992}, which used a simplified variant of the RPE
    \begin{equation}
        R\frac{d^2R}{dt^2}+\frac{3}{2}\left(\frac{dR}{dt}\right)^2=\frac{p_v-p}{\rho_l}\label{eq:locHom:1},
    \end{equation}
    in which the effects of surface tension and viscosity have been neglected.
    Additionally, the authors assume that the bubble number density $n_0$ is constant and uniform across the entire domain and that there is no velocity slip between the two phases, i.e. that the cluster of bubbles are convected with the same velocity as the fluid they are immersed in.
    To obtain an equation of motion the authors simplify the structure of the bubble cluster by assuming that the relative positions of the bubbles do not change and that all bubbles share a common radius $R$.
    This implies that the motion of a bubble cluster is governed by the total velocity potential of the bubble cluster given by
    \begin{equation}
        \sum_{j}\frac{1}{r_j}\frac{dR_j}{dt}R_j^2,
    \end{equation}
    where $r_j$ denotes the distance between the center of the $j$th bubble and the center of the cluster.
    Due to this observation, the authors seek to obtain a description of the temporal derivative of the velocity potential in order to obtain a description of the motion of the cluster of bubbles via the expression
    \begin{equation}
        \frac{d}{dt}\left(\sum_{j}\frac{1}{r_j}\frac{dR_j}{dt}R_j^2\right)+R\frac{d^2R}{dt^2}+\frac{3}{2}\left(\frac{dR}{dt}\right)^2=\frac{p_v-p}{\rho_l}\label{eq:locHom:2}
    \end{equation}
    obtained by adding said derivative to the left hand side of \eqref{eq:locHom:1}.
    The authors first note that the assumption of a uniform bubble number density $n_0$ and common radius $R$ leads to the local approximation of the vapor volume fraction $\alpha_v$ as the total volume of the bubble cluster, yielding the relation $\alpha_v=n_0\frac{4}{3}\pi R^3$ previously stated in \eqref{eq:cavmodel_vaporvolfrac_expr}.
    This local approximation of the vapor volume fraction is quite noteworthy, as it forms the cornerstone of all cavitation models of the category RPE.
    Using their assumption that the relative positions of the bubbles do not change and that all bubbles share a common radius $R_j=R$, the authors conclude that
    \begin{equation}
        \begin{split}
            \frac{d}{dt}\sum_{i}\frac{1}{r_j}\frac{dR}{dt}R^2&=2\pi (\Delta r)^2\left(n_0R^2\frac{d^2R}{dt^2}\right .\\
        &\hspace{-2ex}\left.+\frac{dn_0}{dt}R^2\frac{dR}{dt}+2n_0R\left(\frac{dR}{dt}\right)^2\right),\\
        \nabla\cdot \left(\sum_{i}\frac{1}{r_j}\frac{dR}{dt}R^2\right)&=0,
        \end{split}
    \end{equation}
    from which \eqref{eq:locHom:2} may be rewritten as
    \begin{equation}
        \begin{split}
            \frac{p_v-p}{\rho_l}&=\left(\frac{3}{2}+4\pi(\Delta r)^2n_0R\right)\left(\frac{dR}{dt}\right)^2+\frac{d^2R}{dt^2}\\
        &+(2\pi n_0(\Delta r)^2R)\frac{d^2R}{dt^2}+2\pi(\Delta r)^2\frac{dn_0}{dt}R^2\frac{dR}{dt}.
        \end{split}\label{eq:locHom:3}
    \end{equation}
    To complete the model, the authors note that due to the assumptions of incompressible flow and no velocity slip between the two phases, the temporal derivatives in \eqref{eq:locHom:3} may be replaced with material derivatives $\frac{D}{Dt}=\frac{\partial}{\partial t}+\mathbf{u\cdot\nabla}$, leading to the local homogeneous model
    \begin{equation}
        \begin{split}
            (1&+2\pi n_0(\Delta r)^2R)R\frac{D^2R}{Dt^2}+2\pi(\Delta r)^2\frac{Dn_0}{Dt}R^2\frac{DR}{Dt}\\
            &+\left(\frac{3}{2}+4\pi(\Delta r)^2n_0R\right)\left(\frac{DR}{Dt}\right)^2=\frac{p_v-p}{\rho_l}
        \end{split}
    \end{equation}
    describing the motion of a cluster of cavitation bubbles.
    
    Several of the other models highlighted in this review, including the Bubble Density-Liquid Volume Coupling Model\cite{schnerrPhysicalNumericalModeling2001} and the Vapor Nuclei-Adjusted Model\cite{singhalMathematicalBasisValidation2002} directly reference the idea of using the RPE as introduced by this model as a starting point for their approach to modelling cavitation; as such, this model is a significant milestone in the development in cavitation modelling.
    Curiously, the exact approach used by the present model, i.e. constructing a model for the momentum of a cluster of bubbles, has not been pursued further in the literature, instead focusing mostly on constructing various TEMs in which the vapor volume fraction is used as a description of the approximate location of cavities and their size along with appropriate source terms that dictate the production and destruction of various cavities.
    The approach to modelling cavitation used by the Bubble Cluster Model is very appealing when interpreted physically due to the large number of real-world cases of cavitating flows in which the structure of the cavitating region can qualitatively be described as a cloud of bubbles of approximately equal size, which is exactly the type of structure considered in the Bubble Cluster Model.
    However, any application of this cavitation model must account for its deficiencies, most notably the fact that the Bubble Cluster Model has no way to account for neither the creation of new bubbles nor the destruction of existing bubbles, both of which are a major part of cavitation.
    Additionally, the model assumes that the bubbles in the tracked cluster all share a common radius and that their relative positions do not change.
    Both of these assumptions fail to hold in a variety of circumstances; in fact, Ida \cite{idaMultibubbleCavitationInception2009} has shown that the presence of other, larger bubbles in the immediate vicinity of a bubble can greatly impact the growth of the smaller bubble, even leading to the collapse of the smaller bubble.
    \red{
    Kubota et al.\cite{kubotaNewModellingCavitating1992} implemented their model using their in-house finite-difference solver SACT-III \cite{kubotaNewNumericalSimulation1988} and validate their model by applying it to the task of simulating flow over a hydrofoil using varying conditions, e.g. different angles of attack in both cavitating and non-cavitating conditions.
    The results thus obtained show good agreement with experimental data reported by Izumida et al. \cite{izumidaRelationshipCharacteristicsPartial1980} (e1) and also manage to capture the vortex-shedding phenomena observed by Kubota et al.\cite{kubotaUnsteadyStructureMeasurement1989, kubotaFiniteDifferenceAnalysis1989} (e2) in previous experiments and simulations.
    Grandjean et al. \cite{grandjeanShockPropagationLiquids2012} derived a more general model for the dynamics of a bubble cluster that reduces to the Bubble Cluster Momentum model in a special case, then applied this model towards simulating shock propagation in bubbly flows.
    }

\subsection{The Thermodynamic Variable Table Model}
    The Thermodynamic Variable Table Model, proposed by Ventikos and Tzabiras\cite{ventikosNumericalMethodSimulation2000}, is a model for simulating cavitation using pre-existing information regarding the physical properties of the fluid given information about the current pressure and enthalpy of the fluid.
    For the purposes of demonstration, the authors choose water as their cavitating fluid due to existence of reliable sources for the necessary information regarding the physical properties of water.
    The authors simulate cavitation by numerically solving both the Navier-Stokes equations and the transport equation for the stagnation enthalpy $h$ of the mixture at a fixed Reynolds number $Re=2000$, then using the pressure and the enthalpy to calculate fluid properties such as the mixture density $\rho_m$ and the mixture viscosity $\mu_m$ on a cell-by-cell basis.
    Within each cell, the fluid properties are calculated using a table from the U.K. Committee on the Properties of Steam\cite{unitedkingdomcommitteeonthepropertiesofsteamUKSteamTables1970} containing the values of various properties of a water-vapor mixture at a wide range of possible values for the pressure and enthalpy.
    These properties include the mixture viscosity $\mu_m$, absolute temperature $T$, thermal conductivity $K$, specific heat at constant pressure $c_p$, and specific volume $V_{sp}$, from which the mixture density $\rho_m$ can be calculated as $\rho_m=\frac{1}{V_{sp}}$.
    Because of this, the fluid properties may be viewed as functions of the pressure and the enthalpy, i.e.
    \begin{align*}
        \rho_m&=\rho_m(p,h)=\frac{1}{V_{sp}(p,h)},\\
        \mu_m&=\mu_m(p,h),\\
        T&=T(p,h), \\
        K&=K(p,h), \\
        c_p&=c_p(p,h).
    \end{align*}
    hence the model belongs to the category EOS. 
    \red{
    Ventikos and Tzabiras\cite{ventikosNumericalMethodSimulation2000} implemented this method in their in-house model using a finite volume approach with staggered momentum components.
    The implemented model was validated against experimental data of cavitating flow over a hydrofoil recorded by Avellan et al.\cite{avellanGenerationMechanismDynamics1988} (e3), showing moderate agreement with the trends of suction-side pressure coefficients along the hydrofoil length.
    It has been reported by Li and Yu\cite{liCavitationModelsThermodynamic2021} that the model may have difficulties in dealing with three-dimensional problems.
    }

\subsection{The Lattice-Boltzmann EOS Model}
    Introduced by Banerjee and Saritha\cite{banerjeeNumericalStudyCavitation2015}, the Lattice-Boltzmann EOS Model employs an appropriate EOS along with a lattice-Boltzmann method for simulating both the flow and the development of cavitation, using the distribution of the fluid density as a representation of the location and size of cavities in the fluid.
    The authors consider two separate choices of an EOS for use with their model, referred to as the Shan and Chen Equation of State (SCEOS) and the Peng-Robinson Equation of State (PREOS).
    The SCEOS is given by\cite{shanLatticeBoltzmannModel1993}
    \begin{equation}
        p=\frac{\rho}{3}-4435.2\exp\left(-\frac{400}{\rho}\right),
    \end{equation}
     whilst the PREOS is given by\cite{gongNumericalInvestigationDroplet2012}
    \begin{equation}
        p=\frac{\rho RT}{1-b\rho}-\frac{a\rho\alpha(T)}{1+2b\rho-b^2\rho^2},
    \end{equation}
    where
    \begin{equation}
        \begin{split}
            \alpha(T)=\Bigg(1&+(0.37464+1.54226\omega-0.26992\omega^2)\\
        &\hspace{2ex}\times\left(1-\sqrt{\frac{T}{T_c}}\right)\Bigg)^2,
        \end{split}
    \end{equation}
    $\omega$ is the acentric factor of the fluid and the critical temperature $T_c$ and the critical pressure $p_c$ of the fluid are related to the attraction parameter $a$, the repulsion parameter $b$, and the gas ratio $R$ by the formulas
    \begin{align}
        a&= \frac{0.45724 R^2 T_c^2}{p_c},\\
        b&= \frac{0.45724 R T_c}{p_c}.
    \end{align}
    The authors choose the value $\omega=0.3443$, corresponding to the acentric factor of water and assign the values $a=\frac{2}{29},b=\frac{2}{21}$, and $R=1$.
    
    To validate their model, the authors attempt to simulate the growth of a single bubble submerged in water by solving for the bubble radius growth rate $\Dot{R}$ in the RPE using the local pressure predicted by both the SCEOS and the PREOS and comparing the predicted values to the exact values.
    The predicted bubble growth rates using the SCEOS agree very well with the exact values, but the growth rates predicted by the PREOS show large errors that increase with time.
    The authors attribute this discrepancy to the influence of a smaller computational domain in which the effects of walls are more strongly felt.
    \red{
    This model is implemented using the Lattice-Boltzmann method on a D2Q9 grid and employing the exact difference method.
    The model is validated by simulating the saturated liquid and vapor densities at different temperatures using both the PREOS and the SCEOS, obtaining results in agreement with previous theoretical predictions made by Jain et al.\cite{jainLatticeBoltzmannFramework2009} for the PREOS and Kuzmin\cite{kuzminMultiphaseSimulationsLattice2010} for the SCEOS.
    Saritha and Banerjee\cite{sarithaBubbleDynamicsPressuredriven2020} applied this model for further studies of the dynamics and deformation of bubbles in cavitating flows within a micro-scale channel.
    }

\subsection{The Polytropic Closure Model}
    Denner et al. \cite{dennerModelingAcousticCavitation2020} proposed the Polytropic Closure Model as a method for simulating acoustic cavitation in polytropic gas-liquid systems.
    The authors employ the Navier-Stokes equations along with the Noble-Abel stiffened gas model, which yields a polytropic EOS given by
    \begin{equation}
        \rho=\frac{K(p+\Pi)^\Gamma}{1+bK(p+\Pi)^\Gamma},\label{eq:pcm.1}
    \end{equation}
    where $\Gamma=1/\kappa$ and $\kappa$ is the polytropic exponent, $b$ is the co-volume of the non-condensable gas, $\Pi$ is a pressure constant that accounts for the attraction between molecules, and $K$ is a polytropic constant defined using a reference pressure $p_0$ and a reference density $\rho_0$ as
    \begin{equation}
        K=\frac{\rho_0}{(p_0+\Pi)^\Gamma(1-b\rho_0)}.\label{eq:pcm.2}
    \end{equation}
    \red{Note that in the equations \eqref{eq:pcm.1} and \eqref{eq:pcm.2}, the effects of surface tension and gravity have been neglected.}
    In order to validate their model, the authors apply their model to three test cases which each aim to demonstrate aspects characteristic of acoustic cavitation, using four different types of fluids for each test.
    In their first validation test, the authors simulate the propagation of acoustic waves given a small perturbation to the flow.
    First the authors consider the case of propagation of acoustic waves in single-phase flow where the predicted wavelength and pressure amplitude of the wave are compared to their theoretical values, obtaining agreeable results for each of the four fluids.
    Next, the propagation of acoustic waves in gas-liquid flows is investigated by applying the model to predict the amplitude of the pressure pulses in each phase of the two air-water mixtures considered by the authors at the fluid interface and comparing to the theoretical values.
    As before, the predicted values of the pressure amplitudes greatly agree for both flows considered.
    
    The second validation test considered by the authors aims to demonstrate the model's capacity to predict pressure-driven bubble dynamics properly.
    To this end, the authors apply their model to simulate the collapse-expansion cycle of a single spherical air bubble in water as described by the Gilmore equation over a range of combinations of spatial and temporal resolutions.
    The results thus obtained converge to the theoretical solution given by the Gilmore model\cite{gilmoreGrowthCollapseSpherical1952} for sufficiently small time steps and mesh spacings.
    
    The final test considered by the authors aims to simulate the wall-bounded collapse of an air bubble in water situated at various distances from a solid wall, a process that involves complex pressure-driven interactions.
    The results obtained by the authors clearly depicts both the thin liquid film between the bubble and the wall as well as the high-velocity jet directed towards the wall during the final steps of the bubble collapse.
    Additionally, the authors compare the minimum thickness of the liquid film separating the bubble and the wall predicted by the model for a range of dimensionless initial stand-off distances to a set of experimental measurements of the same values by Reuter and Kaiser\cite{reuterHighspeedFilmthicknessMeasurements2019} (exp4), obtaining predictions with a high coefficient of determination and thus implying the model is capable of accurately predicting this process.
    \red{Denner et al.\cite{dennerModelingAcousticCavitation2020} developed their model on a finite volume approach, using a volume-of-fluid multiphase method. The implementation is not published.
    This model was applied by Denner and Schenke\cite{dennerModelingAcousticEmissions2023} in their study of acoustic emissions and shock formation of cavitation bubbles.
    }

\subsection{The Ginzburg-Landau Potential Model}
    Kunz et al.\cite{kunzPreconditionedNavierStokes2000} introduced the Ginzburg-Landau Potential Model, in which cavitation is simulated based on expressions for the source terms in the transport equation of the liquid volume fraction $\alpha_l$; these source terms are reformulated in terms of the vapor volume fraction $\alpha_v$ as mentioned in section \ref{subsec:approach}.
    The authors employ the split of the source term defined in \eqref{eq:source_term_split}, and use two different approaches for deriving expressions for the two source terms.
    For the evaporation term $\Dot{m}^+$, the authors take the production term from a cavitation model developed by Merkle et al.\cite{merkleComputationalModelingDynamics1998}, who model their source term not in terms of the bubble radius, but instead use dimensional arguments based on the dynamics of large-bubble clusters.
    The source term derived by Merkle et al. is given by
    \begin{equation}
        \Dot{m}^+=C_+\frac{\rho_v(1-\alpha_v)(p_v-p)}{0.5\rho_l U_\infty^2 t_\infty},\quad p < p_v,
    \end{equation}
    where $t_\infty=d/U_\infty$ is the characteristic time scale of fluid motion and $C_+$ is an empirical constant.
    The authors express their condensation term $\Dot{m}^-$ based on simplified potentials obtained from the Ginzburg-Landau theory of superconductivity, using arguments similar to that of Kunz et al. \cite{kunzMultiphaseCFDAnalysis1999}.
    The idea of using these potentials is due to the work of Hohenberg and Halperin\cite{hohenbergTheoryDynamicCritical1977}, who described applications of this theory to a variety of physical systems, including the case of two-phase fluids.
    The condensation term is given by
    \begin{equation}
        \Dot{m}^- = C_-\frac{\rho_v(1-\alpha_v)^2\alpha_v}{t_\infty}, \quad p > p_v,
    \end{equation}
    where $C_-$ is another empirical constant.
    Additionally, the mixture viscosity is taken to be the turbulent eddy viscosity expressed in terms of the turbulent kinetic energy $k$ and the turbulent dissipation rate as
    \begin{equation*}
        \mu_{m,t}=\frac{C_\mu \rho_m k^2}{\varepsilon}.
    \end{equation*}
    \red{
    This model is implemented using the UNCLE framework by  Taylor et al.\cite{taylorUnsteadyThreedimensionalIncompressible1995}, employing the finite volume method with $k-\varepsilon$ turbulence modelling.
    Furthermore, it is implemented as one optional cavitation model in OpenFOAM.
    The authors validate their model by simulating cavitating flow over an ogive body; the predicted values of the pressure coefficients across a range of cavitation numbers agree with experimental data reported by Rouse and McNown\cite{rouseCavitationPressureDistribution1948} (e5).
    This model was recently used by Sikirica et al. \cite{sikiricaCavitationModelCalibration2020}, who proposed a machine learning framework for calibrating the empirical constants in cavitation models.
    Using this model as an example, they applied the calibrated model towards simulating cavitating flow over a five-bladed propeller, showing good agreement with the experimental data.}

\subsection{The Interface Mass and Normal Momentum Model}
    Senocak and Shyy's\cite{senocakInterfacialDynamicsbasedModelling2004} model considers the interface between the liquid phase and the vapor phase, and aims to construct a model of the source terms for the liquid volume fraction $\alpha_l$ in terms of the flow characteristics of said interface in the special case of flows at high Reynolds numbers.
    Proceeding as in Carey\cite{careyLiquidVaporPhase2007}, the authors obtain the following expression for conservation of mass and normal momentum at the bubble interface:
    \begin{equation}
        \rho_l(u_{l,n}-u_{i,n})=\rho_v(u_{v,n}-u_{i,n}),\label{eq:intTrans:1.1}
    \end{equation}
    \begin{align}
        \begin{aligned}
        p_v-p_l&=\sigma\frac{R_1+R_2}{R_1R_2}+2\mu_v\frac{\partial u_{v,n}}{\partial n}-2\mu_l\frac{\partial u_{l,n}}{\partial n}\\
        &\hspace{1.65ex}+\rho_l(u_{l,n}-u_{i,n})^2-\rho_v(u_{v,n}-u_{i,n})^2.
        \end{aligned}\label{eq:intTrans:1.2}
    \end{align}
    The authors note here that thermal effects are unaccounted for in the expressions \eqref{eq:intTrans:1.1}, \eqref{eq:intTrans:1.2}.
    Furthermore, the authors choose to neglect the effects of both viscosity and surface tension due to their focus on flows with high Reynolds numbers, where the effects of viscosity and surface tension on cavitation are negligible.
    By considering the mixture density $\rho_m$, expressed here via the liquid volume fraction $\alpha_l$ as
    \begin{equation}
        \rho_m=\rho_l\alpha_l+\rho_v(1-\alpha_l)\label{eq:intTrans:2}
    \end{equation}
    and assuming that the phase change occurs between the vapor and the mixture phases, the expressions \eqref{eq:intTrans:1.1}, \eqref{eq:intTrans:1.2} reduce to
    \begin{align}
        \begin{aligned}
        \rho_m(u_{m,n}-u_{i,n})&=\rho_v(u_{v,n}-u_{i,n}),
        \end{aligned}\label{eq:intTrans:3.1}\\
        \begin{aligned}
        p_v-p_l&=\rho_m(u_{m,n}-u_{i,n})^2\\
        &\hspace{3ex}-\rho_v(u_{v,n}-u_{i,n})^2.
        \end{aligned}\label{eq:intTrans:3.2}
    \end{align}
    Rewriting \eqref{eq:intTrans:3.1} as
    \begin{equation}
        u_{m,n}-u_{i,n}=\rho_v\frac{u_{v,n}-u_{i,n}}{\rho_m},\label{eq:intTrans:4}
    \end{equation}
    the reduced normal momentum expression \eqref{eq:intTrans:3.2} can be restated via \eqref{eq:intTrans:4} as
    \begin{equation}
        p_v-p_l=\rho_v(u_{v,n}-u_{i,n})^2\left(\frac{\rho_v}{\rho_m}-1\right).\label{eq:intTrans:5}
    \end{equation}
    Recalling the definition of the mixture density \eqref{eq:intTrans:2}, \eqref{eq:intTrans:4} may be rewritten in order to obtain a final expression of the liquid volume fraction $\alpha_l$ in terms of the liquid-vapor interface characteristics:
    \begin{equation}
        \alpha_l= \frac{p_l-p_v}{(u_{v,n}-u_{i,n})^2(\rho_l-\rho_v)}\left(\frac{\rho_l}{\rho_v}\alpha_l+1-\alpha_l\right)\label{eq:intTrans:6}
    \end{equation}
    In order to obtain expressions for the source terms in the transport equation of $\alpha_l$ from the expression \eqref{eq:intTrans:6}, the authors adopt an approach from turbulence modelling described in Wilcox \cite{wilcoxTurbulenceModelingCFD1994}, in which the source terms are derived from normalizing the expression \eqref{eq:intTrans:6} using a characteristic time scale $t_\infty=L_{\text{ch}}/u_\infty$ that is consistent with the definition of the Reynolds number, i.e. the rate of generation $\Dot{S}$ of $\alpha_l$ is expressed as
    \begin{align}
        \begin{aligned}
        \Dot{S}=\frac{\alpha_l}{t_\infty}&=\frac{\rho_l(p_l-p_v)\alpha_l}{\rho_v(u_{v,n}-u_{i,n})^2(\rho_l-\rho_v)t_\infty}\\
        &\hspace{4ex}+\frac{(p_l-p_v)(1-\alpha_l)}{(u_{v,n}-u_{i,n})^2(\rho_l-\rho_v)t_\infty}.
        \end{aligned}\label{eq:intTrans:7}
    \end{align}
    With this in mind, the transport equation of $\alpha_l$ reduces to
    \begin{align}
        \begin{aligned}
            \frac{\partial \alpha_l}{\partial t}\hspace{-.5ex}+\hspace{-.5ex}\nabla\hspace{-.5ex}\cdot(\alpha_l\mathbf{u})&=\frac{\rho_l(p_l-p_v)\alpha_l}{\rho_v(u_{v,n}-u_{i,n})^2(\rho_l-\rho_v)t_\infty}\\
            &\hspace{1ex}+\frac{(p_l-p_v)(1-\alpha_l)}{(u_{v,n}-u_{i,n})^2(\rho_l-\rho_v)t_\infty},
        \end{aligned}\label{eq:intTrans:8}
    \end{align}
    yielding the following expressions for the source terms:
    \begin{align}
        \Dot{m}^+ &= \frac{\rho_l(p_l-p_v)\alpha_l}{\rho_v(u_{v,n}-u_{i,n})^2(\rho_l-\rho_v)t_\infty},\quad \label{eq:intTrans:9.1} \\
        \Dot{m}^- &= \frac{(p_l-p_v)(1-\alpha_l)}{(u_{v,n}-u_{i,n})^2(\rho_l-\rho_v)t_\infty}.\label{eq:intTrans:9.2}
    \end{align}
    The authors note here that as a consequence of choosing a time scale consistent with the Reynolds number, the source terms derived in this model express the mass transfer rate of a cluster of bubbles and not the mass transfer rates of a single bubble.
    \red{Furthermore, the expressions for the source terms in \eqref{eq:intTrans:8} do not contain any empirical factors and are instead formulated in terms of adjustable parameters representing physical factors such as momentum and pressure, in contrast to most other cavitation models.}
    To couple the source terms \eqref{eq:intTrans:9.1} and \eqref{eq:intTrans:9.2} to the equations governing the flow characteristics, the source terms are modified as
    \begin{align}
        \Dot{m}^+ &= \frac{\rho_l(p_v-p)\alpha_l}{\rho_v(u_{v,n}-u_{i,n})^2(\rho_l-\rho_v)t_\infty}, & p&<p_v,\label{eq:intTrans:10.1} \\
        \Dot{m}^- &= \frac{(p-p_v)(1-\alpha_l)}{(u_{v,n}-u_{i,n})^2(\rho_l-\rho_v)t_\infty}, & p&>p_v.\label{eq:intTrans:10.2}
    \end{align}
    
    The authors applied the model to simulate the cavitating flow within two convergent-divergent nozzles with different geometries; due to the different geometries of the nozzles, the flow through the first nozzle forms unsteady cavitation with prominent cloud shedding, while the flow through the second nozzle forms a stable sheet cavity with minimal shedding.
    These nozzles were previously investigated experimentally by Stutz and Reboud\cite{stutzExperimentsUnsteadyCavitation1997,stutzTwophaseFlowStructure1997,stutzMeasurementsUnsteadyCavitation2000} (exp6), who reported time-averaged velocity and vapour volume fraction profiles within the cavity of both nozzles and provided qualitative description of the cavity behaviour previously mentioned.
    The authors performed both steady-state as well as time-dependent simulations of the cavitating flow through the nozzles; the steady-state results were reported in the first part of the article\cite{senocakInterfacialDynamicsbasedModelling2004}, whilst the time-dependent calculations were reported in the second part \cite{senocakInterfacialDynamicsbasedModelling2004a}.
    The results obtained using Interface Mass and Normal Momentum Model to simulate the cavitating flow through both nozzles showed a trend in the computed time-averaged velocity and vapour volume fraction profiles similar to the trend in the experimental data reported by Stutz and Reboud.
    However, the steady-state results overestimated the vapor content towards the end of the cavity in comparison with the experimental data; this discrepancy was remedied in the time-dependent results\cite{senocakInterfacialDynamicsbasedModelling2004a}.
    This model is a departure from using the view of cavitation as the growth and collapse of bubbles as a means of expressing the transport of vapor via the radial growth of said bubbles, instead focusing directly on the mass exchange occurring directly at the interface between the two phases.
    This avoids the problem of approximating a solution of the equation governing the radial growth of the bubbles, i.e. the RPE.
    Furthermore, the choice to use a time scale consistent with the Reynolds number yielding the mass transfer rates of a cluster of bubbles yields a comparison of this model to the Bubble Cluster Momentum Model\cite{kubotaNewModellingCavitating1992}, which expressed the motion of a cluster of bubbles using the RPE to express the radial growth of each bubble in the cluster.
    \red{The model was developed as an in-house tool, using the finite volume method, as well as the $k-\varepsilon$ turbulence model.    
    Utturkar et al. \cite{utturkarComputationalModelingThermodynamic2005} expanded this model, accounting also for thermal effects and applying it to simulate cavitation in cryogenic flows.}

\subsection{The Thermodynamic Interface Model}
    The Thermodynamic Interface Model, introduced by Deshpande et al.\cite{deshpandeNumericalModelingThermodynamic1997}, models the shape of the liquid-vapor interface in sheet cavitating flows with a single, large cavity by approximating the thermal boundary around the cavity.
    This is accomplished by coupling the Navier-Stokes equations with the energy equation and specifying appropriate boundary conditions at the interface for the quantities of interest, namely the pressure, the velocity, and the temperature.
    The authors assume that the cavity is a region of constant pressure below the vapor pressure, with any given point on the cavitating surface classified as either a cavitating point or a solid wall based on the local pressure being above or below the vapor pressure.
    Once the cavity points have been determined, the model then determines the exact location and shape of the cavity from tracing the streamlines of the flow, from which the thickness of the cavity can be determined.
    To ensure a valid profile of the cavity, the authors enforce a strict positiveness of said thickness in order to prevent the cavity from moving inside the surface and apply a specific boundary condition for the energy equation at the liquid-vapor interface given by
    \begin{equation}
        \frac{dT}{dn}=-\frac{\rho_vkH_{fg}}{\rho_l^2 C_p}\frac{dQ}{ds},
    \end{equation}
    where $H_{fg}$ is the heat of vaporization, $k$ is the fluid's thermal conductivity, $C_p$ is the specific heat of the fluid, and $dQ$ is the volume flow rate of the vapor added to the cavity in the area $ds$ of the interface.
    This condition expresses the thermal depression caused at the interface, which forms a thermal boundary around the cavity.
    \red{The authors implemented their model using as a steady-state method, with $4^\text{th}$ order Runge-Kutta pseudo time stepping and a central difference spacial derivative approximation. Turbulence was modelled using the algebraic Baldwin-Lomax method\cite{baldwinThinlayerApproximationAlgebraic1978}.
    The authors validate their model by simulating the flow of water over a hydrofoil, obtaining predicted contours of the pressure whose shapes resemble those observed in experiments by Shen and Dimotakis\cite{shenInfluenceSurfaceCavitation1989} (e7).
    Additionally, the authors investigate the model's capability to simulate thermodynamics effects of cavitating flow of both liquid hydrogen and liquid nitrogen over a two-dimensional airfoil by comparing the predicted temperature depression across the surface of the airfoil with experimental data by Hord\cite{hordCavitationLiquidCryogens1973} (e8), demonstrating a moderate agreement between the predicted values and the experimental data.}
    % \blue{
    % No references to other applications of this model were found in the literature; one possible explanation may be the model's difficulties in dealing with three-dimensional problems as noted in the review of Li and Yu\cite{liCavitationModelsThermodynamic2021}.
    % }

\subsection{The Wake Closure Model}
    Introduced by Liu et al. \cite{liuNumericalMethodSimulation2006}, the Wake Closure Model attempts to determine the shape of the liquid-vapor interface in cavitating flows by using the Reynolds-averaged Navier-Stokes equations to simulate the turbulent flow along with an appropriate model for the turbulent viscosity $\mu_t$ that accounts for the fact that the area most dominated by turbulence in cavitating flows is the wake downstream of an attached cavity, where vapor bubbles collapse suddenly and strongly interact with any solid wall.
    To this end, the authors employ the Baldwin-Lomax turbulence model\cite{baldwinThinlayerApproximationAlgebraic1978} to estimate $\mu_t$ locally.
    For any point located at a normal distance $y$ from a solid wall, this model defines $\mu_t$ as follows:
    \begin{equation}
        \mu_t=\begin{cases}
        0.16\rho y^2\left(1-\text{e}^{-y^+/A^+}\right)^2|\Omega|, & 0\leq y\leq y',\\
        0.02688\rho F_w\left(1+5.5\left(\frac{0.3y}{y_{\max}}\right)^6\right)^{-1}, & y\geq y',
        \end{cases}
    \end{equation}
    where $y'$ is an adjustable parameter defining the separation of the flow into an inner and outer layer, $A^+=26$, $y^+$ is defined using the shear stress $\tau_w$, fluid density $\rho_w$ and molecular viscosity $\mu_w$ at the wall as $y^+=\frac{\sqrt{\rho_w\tau_w}}{\mu_w}$, $\Omega$ is the strength of the vortex in the flow, $y_{\max}$ is the maximum point of the function
    \begin{equation}
        F(y)=y|\Omega|\left(1-\text{e}^{-y^+/A^+}\right)
    \end{equation}
    with corresponding maximal value $F_{\max}$, and $F_w$ is defined using the maximal $\mathbf{u}_{\max}$ and minimal velocity $\mathbf{u}_{\min}$ of the flow as
    \begin{equation}
        F_w=\min\left\{\frac{0.25\|\mathbf{u}_{\max}-\mathbf{u}_{\min}\|^2}{F_{\max}},y_{\max}F_{\max}\right\}.
    \end{equation}
    Assuming that the cavity surface is a free surface and that the pressure inside the cavity and on the boundary of the cavity is constant and equal to the vapor pressure, the model approximates the shape of the surface of the cavity along the length of the solid wall by first searching for the first point along the wall at which the local pressure is minimal and less than the vapor pressure, then declaring each point downstream of said point to belong to the cavity if the local pressure is below the vapor pressure.
    After recalculating the flow properties with the free surface condition enforced along the surface of the cavity, the model checks if the pressure distribution along the surface is approximately equal to the vapor pressure and then iteratively adjusts the local thickness $r(s)$ of the cavity at every point $s$ along its entire length until this condition is satisfied.
    The updated local thickness $r^{(n+1)}$ above a point is defined from the current local thickness $r^{(n)}$ and a relaxation coefficient $\lambda$ with $|\lambda|\leq 1$ as $r^{(n+1)}=r^{(n)}+\lambda\Delta r^{(n)}$, with the adjustment $\Delta r^{(n)}$ defined using the local pressure difference $p^{(n)}-p_v$ and the pressure gradient $\frac{\partial p^{(n)}}{\partial s}$ along the wall at the current iteration along with the initial local pressure difference $p^{(0)}-p_v$ and pressure gradient $\frac{\partial p^{(0)}}{\partial s}$ as
    \begin{equation}
        \begin{split}
            \Delta r^{(n)}&=\frac{\pi}{180}\int_{s_b}^{s} \beta_0\operatorname{sign}(p_v-p^{n})\sqrt{\frac{|p_v-p^{n}|}{\|p_v-p^{0}\|}}\\
            &\hspace{6ex}+\beta_1\operatorname{sign}\left(\frac{\partial p^{(n)}}{\partial s}\right)\sqrt{\frac{\left|\frac{\partial p^{(n)}}{\partial s}\right|}{\left\|\frac{\partial p^{(0)}}{\partial s}\right\|}}\,\text{d}s,
        \end{split}
    \end{equation}
    where
    \begin{align*}
        \|p_v-p^{0}\|&=\sqrt{\frac{\int_{s_b}^{s_e} (p_v-p^{0})^2\,\text{d}s}{s_e-s_b}},\\
        \left\|\frac{\partial p^{(0)}}{\partial s}\right\|&=\sqrt{\frac{\int_{s_b}^{s_e} \left(\frac{\partial p^{(0)}}{\partial s}\right)^2\,\text{d}s}{s_e-s_b}},
    \end{align*}
    and $s_b$ and $s_e$ denote the inception point and the endpoint of the cavity.
    Once convergence in this iterative procedure has been achieved, the model redefines the surface of the cavity using both the previously determined endpoints of the cavity $s_b$ and $s_e$ as well as a new point $s_w$, which denotes the beginning of the wake region.
    The new point $s_w$ is defined as the first point on the surface after $s_b$ at which the local thickness $r(s_w)$ is decreasing and less than half of the maximal local thickness.
    Using the three points $s_b$, $s_w$, and $s_e$, the cavity surface is defined as the cubic Hermite polynomial interpolating between the points $s_b$ and $s_e$ on the solid wall as well as the point $(s_w,r(s_w))$.
    This approach to capturing the surface of the cavity was chosen because of its efficiency and due to the lack of model available at the time capable of accounting for the turbulence in the wake region and the violation of the condition of constant vapor pressure along the surface of the cavity.
    \red{This model is implemented as an in-house tool, using the SIMPLEC momentum-pressure coupling and the Baldwin-Lomax turbulence model\cite{baldwinThinlayerApproximationAlgebraic1978}.
    The model was validated by simulating the pressure distributions of cavitaing flow over an ogival headform, indicating good agreement with data from  Rouse and McNown\cite{rouseCavitationPressureDistribution1948} (e5).
    Liu et al.\cite{liNumericalPredictionHydrodynamic2007} applied this model for simulating sheet cavitation on a cylindrical headform, showing good agreement with experimental data.
    No other references to other applications of this model were found in the literature; one possible explanation may be the model's difficulties in dealing with three-dimensional problems as noted in the review of Li and Yu\cite{liCavitationModelsThermodynamic2021}.
    }
    % Ishikawa et al.\cite{ishikawaNumericalAnalysisUnsteady2014} investigated the possibility of developing an interface tracking model with the capability of simulating both sheet cavities as well as other types of cavitation.
    %     Their model was validated by simulating both cloud cavitation and supercavitation over a hydrofoil, and the results obtained by the model agree with previous experimental data

\subsection{The Bubble Density-Liquid Volume Coupling Model}
    Schnerr and Sauer\cite{schnerrPhysicalNumericalModeling2001} present a TEM which expresses the source terms using the local bubble radius $R$ and the bubble density $n_0$, assumed uniform throughout the domain.
    Instead of simply using the relation \eqref{eq:cavmodel_vaporvolfrac_expr}, the authors seek an expression for the vapor volume fraction $\alpha_v$ that directly couples the density of vapor bubbles and the liquid volume, corresponding to the physical observation that an increase in the density of bubbles should correspond to a decrease in the liquid volume fraction.
    The rationale given by the authors for their search for a new expression for $\alpha_v$ is the fact that the standard volume-of-fluid method is capable of accounting for convective transport of the vapor volume fraction, but not the change in volume fraction due to phase transition.
    To this end, the authors define the vapor volume fraction $\alpha_v$ for each cell in the computational grid as the ratio of the volume occupied by the vapor within a given cell, denoted $V_v$, and the total volume of said cell, denoted $V_{cell}$.
    Letting $V_l=V_{cell}-V_v$ denote the volume occupied by the liquid within the cell, the authors approximate $V_v$ using the relation \eqref{eq:cavmodel_vaporvolfrac_expr} instead, i.e. $V_v$ is approximated in terms of the bubble number $n_0$ and bubble radius $R$ as
    \begin{equation}
        V_v=\frac{4}{3}n_0\pi R^3.\label{eq:SSM:1}
    \end{equation}
    Recalling that $V_{cell}=V_l+V_v$, the approximation of $V_v$ given by \eqref{eq:SSM:1} implies that the vapor volume fraction can be approximated as
    \begin{equation}
        \alpha_v=\frac{\frac{4}{3}n_0\pi R^3}{1+\frac{4}{3}n_0\pi R^3}.\label{eq:SSM:2}
    \end{equation}
    With the expression \eqref{eq:SSM:2}, the transport equation for $\alpha_v$ may be restated as
    \begin{equation}
        \frac{\partial \alpha_v}{\partial t} + \nabla\cdot (\alpha_v\mathbf{u})=\frac{\frac{4}{3}n_0\pi }{1+\frac{4}{3}n_0\pi R^3}\frac{d}{dt}(R^3);\label{eq:SSM:3}
    \end{equation}
    here, the authors note that \eqref{eq:SSM:3} directly couples the bubble density as expressed by $n_0$ with the liquid volume.
    In order to complete their expression for the source term of $\alpha_v$ in \eqref{eq:SSM:3}, the authors approximate the rate of change of the bubble radius $\frac{dR}{dt}$ as
    \begin{equation}
        \frac{dR}{dt}=\sqrt{\frac{2|p_v-p|}{3\rho_l}}.\label{eq:SSM:4}
    \end{equation}
    This expression is obtained from the RPE \eqref{eq:rayleighPlesset} with the effects of the second-order terms, viscosity, and surface tension all neglected, the pressure at the liquid-bubble interface taken to be equal to the vapor pressure $p_v$, and the reference pressure $p_\infty$ taken to be equal to the cell pressure $p$.
    Combining \eqref{eq:SSM:3} and \eqref{eq:SSM:4} yields the final expressions for the source terms
    \begin{align}
        \Dot{m}^+ &= \frac{4n_0\pi R^2}{1+\frac{4}{3}n_0\pi R^3}\sqrt{\frac{2(p_v-p)}{3\rho_l}}, & p&<p_v,\label{eq:SSM:5.1}\\
        \Dot{m}^- &= \frac{4n_0\pi R^2}{1+\frac{4}{3}n_0\pi R^3}\sqrt{\frac{2(p-p_v)}{3\rho_l}}, & p&>p_v.\label{eq:SSM:5.2}
    \end{align}
    Notably, no empirical constants besides the uniform bubble number $n_0$ are used in the expressions \eqref{eq:SSM:5.1} and \eqref{eq:SSM:5.2}.
    \red{
    The Schnerr and Sauer model is implemented in a volume-of-fluid framework. 
    It is the default model for cavitating flow modelling in Ansys Fluent and an optional cavitation model in OpenFOAM and STAR-CCM+.
    The model was validated through simulation of cavitating flow over a hydrofoil under varying conditions, showing good agreement with previous observations made by Keller and Arndt\cite{kellerCavitationScaleEffects2001} (e9). 
    This model was recently applied by Moganaradjou et al.\cite{moganaradjouEffectSecondaryPassages2023} for simulation of cavitation in a rocket pump.
    }
    
\subsection{The Full Cavitation Model}
    Singhal et al.\cite{singhalMathematicalBasisValidation2002} introduced a cavitation model also based on viewing cavitation as the phenomenon of existing vapor nuclei in the liquid beginning to grow once the local pressure decreases below the saturated vapor pressure of the liquid, then shrinking once the local pressure rises above the vapor pressure.
    The present model considers the governing equation of the vapor mass fraction, denoted $f_v$, and seeks to construct expressions for the evaporation term $\Dot{m}^+$ and the condensation term $\Dot{m}^-$.
    The authors accomplish this goal using the above mentioned viewpoint to first restate the transport equation of the vapor mass fraction in terms of the mixture density and the vapor volume fraction, then reformulating the vapor volume fraction using the assumption of a uniform number of vapor nuclei present in the liquid with a uniform radius as well as \eqref{eq:cavmodel_vaporvolfrac_expr}.
    The authors incorporate the effects of bubble dynamics into their model by using the RPE \eqref{eq:rayleighPlesset} to model the growth of a single bubble's radius, neglecting the effects of interaction between two distinct bubbles. 
    Furthermore, the authors simplify the RPE via neglecting the effects of viscosity, surface tension, and the second-order derivative of the bubble radius, justifying the latter exclusion as the second-order derivative being "important mainly during initial bubble acceleration". 
    Restating their expressions in terms of $f_v$, this yields the initial model
    \begin{align}
        \begin{aligned}
            \Dot{m}^+ &= (4n_0\pi)^{1/3}(3\alpha_v)^{2/3}\frac{\rho_v\rho_l}{\rho_m}\sqrt{\frac{2}{3}\frac{p_b-p_\infty}{\rho_l}}, \\
            \Dot{m}^- &= (4n_0\pi)^{1/3}(3\alpha_v)^{2/3}\frac{\rho_v\rho_l}{\rho_m}\sqrt{\frac{2}{3}\frac{p_\infty-p_b}{\rho_l}}.
        \end{aligned}\label{eq:singhal_1}
    \end{align}
    In order to remove the bubble number density as a parameter of the model, the authors restate their source terms using a method introduced in the nuclear industry cf. Markatos and Singhal\cite{markatosNumericalAnalysisOnedimensional1982}.
    This method forms a correlation between the bubble radius and the local relative velocity between the two phases as well as the surface tension, thus re-introducing surface tension as an effect into their model.
    Using this correlation along with several limiting arguments as $\alpha_v\rightarrow 0$ and an assumption that the phase change rates are proportional to the local relative velocity, the models in \eqref{eq:singhal_1} can be rewritten as
    \begin{align}
        \begin{aligned}
            \Dot{m}^+ &= C_+\frac{u_{ch}}{\sigma}\rho_l\rho_v\sqrt{\frac{2}{3}\frac{p_b-p_\infty }{\rho_l}}(1-f_v), \\
            \Dot{m}^- &= C_-\frac{u_{ch}}{\sigma}\rho_l\rho_l\sqrt{\frac{2}{3}\frac{p_\infty -p_b}{\rho_l}}f_v,
        \end{aligned}\label{eq:singhal_2}
    \end{align}
    where $C_+$ and $C_-$ are empirical constants and $u_{ch}$ is a characteristic velocity that reflects the impact of the local relative velocity between the liquid phase and the vapor phase on the phase change rates.
    To account for the effect of turbulence on cavitation, two further assumptions are made. First, the local relative velocity is eliminated as a parameter of the model through the assumption that the local relative velocity is proportional to the square root of the local turbulent kinetic energy of the flow.
    Secondly, the fluctuations of pressure in turbulent flow are accounted for in the model by first taking the bubble pressure $p_b$ in the expressions \eqref{eq:singhal_1} and \eqref{eq:singhal_2} to be equal to a modified vapor pressure $\Tilde{p}_v$ given by
    \begin{equation}
    \Tilde{p}_v=p_{v}+\frac{0.39\rho_m k}{2},
    \end{equation}
    in accordance with Hinze\cite{hinzeTurbulenceIntroductionIts1975}.
    Replacing the ambient pressure $p_\infty$ in the models \eqref{eq:singhal_2} with the local pressure $p$, the assumptions stated above leads to \eqref{eq:singhal_2} being restated as
    \begin{equation}
        \begin{split}
            \Dot{m}^+ &= C_+\frac{\sqrt{k}}{\sigma}\rho_l\rho_v\sqrt{\frac{2}{3}\frac{\Tilde{p}_v-p}{\rho_l}}(1-f_v),& p &< \Tilde{p}_v \\
            \Dot{m}^- &= C_-\frac{\sqrt{k}}{\sigma}\rho_l\rho_l\sqrt{\frac{2}{3}\frac{p-\Tilde{p}_v}{\rho_l}}f_v, & p &> \Tilde{p}_v.\label{eq:singhal_3}
        \end{split}
    \end{equation}
    The final effect accounted for in this model, that of non-condensable gas, is introduced via the assumption that the cavitating fluid also contains a finite amount of non-condensable gas in dissolved state, i.e. a three-phase flow.
    Additionally, the mass fraction of the non-condensable gas, denoted $f_g$, is assumed to be constant and specified as an input parameter of the model.
    The impact of this assumption on the models \eqref{eq:singhal_3} is that the liquid mass fraction $1-f_v$ is modified as $1-f_v-f_g$, leading to the final models for the phase change rates given below:
    \begin{equation}
        \begin{split}
            \Dot{m}^+ &= C_+\frac{\rho_l\rho_v}{\sigma}\sqrt{\frac{2k}{3}\frac{\Tilde{p}_v-p}{\rho_l}}(1-f_v-f_g), &p &< \Tilde{p}_v,\\
            \Dot{m}^- &= C_-\frac{\rho_l\rho_l}{\sigma}\sqrt{\frac{2k}{3}\frac{p-\Tilde{p}_v}{\rho_l}}f_v, & p &> \Tilde{p}_v.\label{eq:singhal_4}
        \end{split}
    \end{equation}
    The authors of the model have specified recommended values for the empirical constants $C_+$ and $C_-$ based on various validation tests, the results of which are not disclosed in the original article.
    \red{Singhal et al. implement their model using the finite volume method with a standard $k-\varepsilon$ turbulence model. This model is one of the optional cavitation solvers in Ansys Fluent.
    The model was validated by simulating cavitating flow over a conical head as well as a hydrofoil, showing good agreement with data reported by Rouse and McNown\cite{rouseCavitationPressureDistribution1948} (e5) in the former case and data reported by Shen and Dimotakis\cite{shenInfluenceSurfaceCavitation1989} (e7) in the latter case.
    This model was recently applied by Yuan et al.\cite{yuanTheoreticalModelDynamic2022} in their development of a model for dynamic bulk modulus for aerated hydraulic fluids.
    }

\subsection{The Vapor Nuclei-Adjusted Model}
    Zwart et al.\cite{zwartTwophaseFlowModel2004} present a TEM in which source terms are derived by combining the Rayleigh equation, i.e. the RPE \eqref{eq:rayleighPlesset} with the effects of radial acceleration, viscosity, and surface tension all neglected, along with the expression \eqref{eq:cavmodel_vaporvolfrac_expr}, yielding an initial expression of the total mass transfer rate of a single bubble during bubble growth as
    \begin{equation}
        \Dot{m} = \frac{3\alpha_v\rho_v}{R}\sqrt{\frac{p_v-p}{\rho_l}}.\label{eq:ZGB:1}
    \end{equation}
    The expression \eqref{eq:ZGB:1} is immediately generalized to an expression for the total mass transfer rate of a single bubble during bubble collapse as
     \begin{equation}
         \Dot{m} = C\frac{3\alpha_v\rho_v}{R}\sqrt{\frac{|p_v-p|}{\rho_l}}\text{sign}(p_v-p),\label{eq:ZGB:2}
    \end{equation}
    where $C$ is an empirical constant used for calibrating the model.
    The authors note here that the expression \eqref{eq:ZGB:2} works well during bubble collapse, but is both physically incorrect and numerically unstable during bubble growth.
    To remedy this, the authors seek to modify the mass transfer rate to account for the interaction between distinct bubbles. 
    Noting that as an increase in the vapor volume fraction $\alpha_v$ corresponds to a decrease in the density of vapor nuclei, the authors modify the mass transfer rate \eqref{eq:ZGB:2} by replacing $\alpha_v$ with the expression $\alpha_{nuc}(1-\alpha_v)$ during bubble growth, i.e. when $p<p_v$, where $\alpha_{nuc}$ is a parameter of the model specifying the vapor nuclei volume fraction.
    This leads to the final expressions for the evaporation and condensation terms given by
    \begin{equation}
        \begin{split}
            \Dot{m}^+ &= C_{+}\frac{3\alpha_{nuc}(1-\alpha_v)\rho_v}{R}\sqrt{\frac{p_v-p}{\rho_l}},& p&<p_v,\\
        \Dot{m}^- &= C_-\frac{3\alpha_v\rho_v}{R}\sqrt{\frac{p-p_v}{\rho_l}},& p&>p_v.
        \end{split}
    \end{equation}
    Additionally, the authors observed that the model fails to properly predict the oscillating behaviour of certain unsteady cavitating flow when standard turbulence models were employed.
    To remedy this deficiency, the authors approach suggested by Coutier-Delgosha et al.\cite{coutier-delgoshaEvaluationTurbulenceModel2001} that decreases the effect of turbulent viscosity in cavitating regions.
    Using this approach, the standard expression for the eddy viscosity of the mixture
    \begin{equation*}
        \mu_{m,t}=\frac{C_\mu \rho_m k^2}{\varepsilon}
    \end{equation*}
    is modified as
    \begin{equation*}
        \mu_{m,t}=\left(\rho_v+\left(\frac{\rho_v-\rho_m}{\rho_v-\rho_l}\right)^P (\rho_l-\rho_v)\right)C_\mu\frac{k^2}{\varepsilon},
    \end{equation*}
    where $P>1$ is an empirical parameter.
    \red{This method is an optional cavitation method in Ansys Fluent. It is implemented using the finite volume method and the standard $k-\varepsilon$ model.
    The authors validate their model by simulating a variety of cavitating flow conditions previously reported on by different authors: the pressure profile of cavitating flow over a hydrofoil investigated by Shen and Dimotakis\cite{shenInfluenceSurfaceCavitation1989} (e7), cavitating flow in an inducer investigated by Bakir et al.\cite{bakirNumericalExperimentalInvestigations2004} (e10), and cavitating flow in a Venturi nozzle investigated by Stutz and Reboud\cite{stutzExperimentsUnsteadyCavitation1997} (e6).
    In all cases, the model shows good agreement with the reported data.
    This model was applied by Zhou et al. \cite{zhouStudyThermalProperties2023} to investigate thermal properties of oil-film viscosity in squeeze film dampers.}
    
\subsection{The Viscosity-Oriented Model}
    Konstantinov et al.\cite{konstantinovNumericalCavitationModel2014} develop a TEM with the stated intent of obtaining a model that more accurately captures the effects of bubble dynamics on cavitation without sacrificing numerical stability.
    The authors choose the source terns from the Vapor Nuclei-Adjusted Model\cite{zwartTwophaseFlowModel2004} given by
    \begin{equation}
        \begin{split}
            \Dot{m}^+ &= C_{+}\frac{3\alpha_{nuc}(1-\alpha_v)\rho_v}{R}\sqrt{\frac{p_v-p}{\rho_l}},& p&<p_v, \\
        \Dot{m}^- &= C_-\frac{3\alpha_v\rho_v}{R}\sqrt{\frac{p-p_v}{\rho_l}},& p&>p_v
        \end{split}\label{eq:visModel:1}
    \end{equation}
    as a starting point, and model the bubble dynamics using a non-dimensional RPE
    \begin{equation}
            \overline{R}\:\ddot{\overline{R}}+\frac{3}{2}\dot{\overline{R}^2}+\frac{1}{\overline{R}}\frac{1}{\mathrm{Re}}\dot{\overline{R}}=1\label{eq:visModel:2}
        \end{equation}
    with nondimensional variables and numbers
    \begin{equation}
    \overline{R}=\frac{R}{R_0}, \,\, \mathrm{Re}=\frac{R_0\sqrt{|p_v-p|\rho_l}}{4\mu},\,\, \tau=\frac{t}{R_0}\sqrt{\frac{|p_v-p|}{\rho_l}}.
    \end{equation}
    The authors seek to express the bubble growth rate in terms of the Reynolds number $\text{Re}$, neglecting the effects of surface tension and elected to use the non-corrected expression for the eddy viscosity given by
    \begin{equation*}
        \mu_{m,t}=\frac{C_\mu \rho_m k^2}{\varepsilon}.
    \end{equation*}
    Additionally, the model assumes that both the liquid and vaporous phase are incompressible in order to facilitate simpler expressions.
    Through a series of numerical calculations based on \eqref{eq:visModel:2}, the authors report the new expression
    \begin{align}
        \begin{aligned}
        \frac{dR}{dt}&=\tanh \left[1.221\left(\frac{R_0\sqrt{|p_v-p|\rho_l}}{4\mu}\right)^{0.353}\right]\\
        &\hspace{4ex}\times\sqrt{\frac{2|p_v-p|}{3\rho_l}}.
        \end{aligned}\label{eq:visModel:3}
    \end{align}
    Replacing the approximation $\frac{dR}{dt}=\sqrt{\frac{2|p_v-p|}{3\rho_l}}$ in the source terms \eqref{eq:visModel:1} from the Vapor Nuclei-Adjusted Model\cite{zwartTwophaseFlowModel2004} with the right hand side of the expression \eqref{eq:visModel:3}, denoted by $\Tilde{R}(p)$, the authors obtain the new source terms
    \begin{align}
        \Dot{m}^+ &= C_{+}\frac{3\alpha_{nuc}(1-\alpha_v)\rho_v}{R}\Tilde{R}(p), & p&<p_v,\label{eq:visModel:4.1} \\
        \Dot{m}^- &= C_-\frac{3\alpha_v\rho_v}{R}\Tilde{R}(p), & p&>p_v.\label{eq:visModel:4.2}
    \end{align}
    \red{This model was developed in the finite element solver Ansys CFX, using the standard $k-\varepsilon$ turbulence model.
    The model was validated against experimental data on cavitating flow in a "pipe-pipe" jet element, obtaining results in good agreement with the experimental data as well as previous observations made by Ackeret\cite{ackeretExperimentelleUndTheoretische1930} (e11).
    This model was applied by Konstantinov et al. \cite{konstantinovAnalyticalCalculationHydraulic2017} towards simulating cavitating flow within a  jet-cavitation fluid mass flow stabilizer.
    }

\subsection{The Plane Surface Evaporation Model}
    In Saito et al.'s \cite{saitoNumericalAnalysisUnsteady2003} model, the source terms are derived using the theory of vaporization and condensation on a plane surface as described by Sone and Sugimoto\cite{soneStrongEvaporationPlane1990}.
    Due to the different theoretical approach, the source terms expressed in this model do not express the mass volume change, but rather the mass surface change.
    Furthermore, this approach also introduces a new parameter of interest used to express the source terms, namely the interfacial area concentration in the liquid-vapor mixture $A$ given by
    $A=C_a\alpha_v(1-\alpha_v)$.
    The source terms of this model are given by
    \begin{align}
        \Dot{m}^+ &= C_+\frac{ A \alpha_v(1-\alpha_v)(p_v- p)}{\sqrt{2\pi R_1 T_{sat}}},& p&<p_v, \\
        \Dot{m}^- &= C_-\frac{\rho_l}{\rho_v}\frac{ A\alpha_v(1-\alpha_v)(p-p_v)}{\sqrt{2\pi R_1 T_{sat}}}, & p&>p_v,
    \end{align}
    where $C_+$, $C_-$, and $C_a$ are empirical constants. The authors state that the relation $C=C_+ C_a=C_- C_a$ holds, but do not elaborate on the exact nature of said relationship.
    Additionally, the authors use an EOS derived for a locally homogeneous gas-liquid mixture to determine the mixture density.
    This EOS was first derived by Okuda and Ikohagi\cite{okudaNumericalSimulationCollapsing1996} and is given in terms of the vapor mass fraction $f_v$, the pressure $p$ and the temperature $T$ as
    \begin{equation}
        \rho_m=\frac{p(p+p_c)}{K(1-f_v)p(T+T_c)+Rf_v(p+p_c)T},
    \end{equation}
    where $p_c$, $T_c$, $K$, and $R$ are the pressure, temperature, liquid, and gas constant of the fluid, respectively.
    Despite the different approach, the expressions source terms of this model are of a similar form to the source terms obtained in other models such as the Vapor Nuclei-Adjusted Model\cite{zwartTwophaseFlowModel2004} and the Full Cavitation Model\cite{singhalMathematicalBasisValidation2002}.
    
    \red{The model was implemented using the finite-volume discretization with cell-centered momentum components. Turbulence was modelled using the Baldwin-Lomax model\cite{baldwinThinlayerApproximationAlgebraic1978}.
    The authors validate their model by simulating cavitating flow over a three-dimensional cylinder; the predicted values of the pressure coefficients across a range of cavitation numbers agree with experimental data reported by Rouse and McNown\cite{rouseCavitationPressureDistribution1948} (e5).
    This model was applied by Mostafa et al.\cite{mostafaNumericalPredictionUnsteady2016} in their study of cavitating flow over a hydrofoil, showing good predicted values for the drag and lift coefficient at various cavitation numbers.
    }

\subsection{The Four-Equation Model}
    Goncalv\`es and Charri\`ere\cite{goncalvesModellingIsothermalCavitation2014} developed a model, which combines the Reynolds-averaged Navier-Stokes equations with expressions for source terms for transport of the vapor volume fraction $\alpha_v$ using two quantities not previously considered in the literature, namely the local speed of sound $c$ and the propagation of acoustic waves without mass transfer $c_{wallis}$.
    As the name of the model implies, the authors choose a set of four equations as the starting point for this model, namely three conservation laws for the mass, momentum, and total energy along with a transport equation for the vapor volume fraction.
    This model is itself derived from previous work of Goncalv\`es \cite{goncalvesNumericalStudyExpansion2013}.
    The system is closed by relating the pressure to the density via a barotropic law introduced by Dellanoy and Kueny\cite{dellanoyTwoPhaseFlow1990}
    \begin{equation}
        \begin{split}
            p(\rho_m)=p_v&+\frac{\left(\rho_l-\rho_v\right)c_{baro}^2}{2}\\
            &\hspace{2ex}\times\arcsin\left(\frac{2\rho_m-\rho_l-\rho_v}{\rho_l-\rho_v}\right),
        \end{split}\label{eq:4eqModel.1}
    \end{equation}
    where $c_{baro}$ is a empirical parameter representing the minimal speed of sound in the mixture region of the fluid.
    The importance of this parameter and other aspects of this EOS were investigated by Goncalv\`es and Patella\cite{goncalvesNumericalSimulationCavitating2009}.
    From \eqref{eq:4eqModel.1}, the speed of sound of the mixture may be be calculated as
    \begin{equation}
        c^2=\left(\frac{\partial p}{\partial \rho_m}\right)_s=\frac{c_{baro}^2}{2\sqrt{\alpha_v(1-\alpha_v)}}
    \end{equation}
    The new effects $c$ and $c_{wallis}$ are related to $\alpha_v$ via two assumptions. First, it is assumed that the total mass transfer rate $\Dot{m}$ is proportional to the divergence $\nabla\cdot\mathbf{u}$ of the velocity field. Proceeding as in Goncalv\`es' previous work, the authors conclude that the proportional relation between $\Dot{m}$ and $\nabla\cdot \mathbf{u}$ is given by
    \begin{equation}
        \Dot{m}=\frac{\rho_l\rho_v}{\rho_l-\rho_v}\left(1-\frac{c^2}{c_{wallis}^2}\right)\nabla\cdot\mathbf{u}.
    \end{equation}
    Secondly, it is assumed that $c_{wallis}$ equals the weighted harmonic mean of the local speed of sound each of the two phases, i.e. that the relation
    \begin{equation}
        \frac{1}{\rho c_{wallis}^2}=\frac{\alpha_v}{\rho_v c_v^2}+\frac{1-\alpha_v}{\rho_l c_l^2}
    \end{equation}
    holds.
    \red{The Four-Equation Model is implemented using a cell-centred finite volume discretization and a matrix-free, implicit time integration method due to Luo et al. \cite{luoFastMatrixfreeImplicit1998}. Turbulence is modelled using the Spalart-Allmaras model\cite{spalartOneEquationTurbulenceModel1992}.
    The model was validated against experimental data of cavitating flow in a Venturi nozzle reported by Barre et al.\cite{barreExperimentsModelingCavitating2009} (e12) and Patella et al.\cite{patellaExperimentsModellingCavitating2006} (e13), showing good agreement.
    The four equation model was used by Goncalv\`es\cite{goncalvesNumericalSimulationCavitating2017} in a comparative study of various turbulence and cavitation models, showing good performance when applied to the test case of water-gas flow in an expansion tube.
    }

\subsection{The Ghost-Fluid Multiscale Model}
    Introduced by Hisao et al.\cite{hsiaoMultiscaleTwophaseFlow2017}, the Ghost-Fluid Multiscale Model is a multiscale model based on the Euler-Lagrangian approach, i.e the model employs the Eulerian approach for simulating the growth and collapse of cavities on the macroscale and tracks the growth and momentum of bubbles on the microscale in a Lagrangian framework.
    The main distinguishing feature of this model is the implementation of the Ghost Fluid Method, a level set method first described by Fedkiw et al. \cite{fedkiwNonoscillatoryEulerianApproach1999} and Kang et al. \cite{kangBoundaryConditionCapturing2000}, to simulate larger deformations in larger cavities such as e.g. sheet cavities as well as a different transition scheme employed by the model to determine of a given cavity currently tracked at the macroscale should switch to the microscale and vice versa.
    Recalling the generic description mentioned at the start of this section, the schemes comprising the Ghost-Fluid Multiscale Model can be described as follows:
    \begin{enumerate}
        \item The macro-scale cavities are resolved in an Eulerian framework by first obtaining the flow characteristics through the Navier-Stokes equations\eqref{eq:NSE}, which are solved using a finite volume method.
        Once the flow characteristics have been resolved, the model approximates the vapor volume fraction locally within each computational cell as a weighted sum of the amount of volume occupied by all cavities currently tracked by the model which at least partly occupy the given cell.
        The weights in this sum are calculated using the same scheme as that of Ma et al. \cite{maEulerLagrangeSimulationsBubble2015}, who assumed that the distribution of the bubbles is approximately a Gaussian distribution centered at the cell of the center with a prescribed standard deviation $R_s$, also referred to as the characteristic spreading radius.
        Having developed the vapor volume fraction $\alpha_{v,i}$ on a per-cell basis, the mixture density $\rho_m$ is then updated in each cell using \eqref{eq:mixtureRhoMu}.
        The final step of the macro-scale scheme is to update the location of existing macro-scale cavities.
        This is done using the Ghost Fluid Method, which identifies the liquid-vapor interfaces as the zero level set of the smooth function $\varphi$ and tracking its evolution via the transport equation
        \begin{equation}
            \frac{D\varphi}{Dt}=\frac{\partial \varphi}{\partial t}+\mathbf{u}_i\cdot\nabla \varphi =0\label{eq:GFMM:phieq1}
        \end{equation}
        coupled with the boundary conditions
        \begin{align}
            \begin{aligned}
                \rho_lp&=\mathbf{n}\cdot(\boldsymbol{\tau}\cdot\mathbf{n})+gz+\sigma\kappa,&\mathbf{n}\cdot(\boldsymbol{\tau}\cdot\mathbf{t})&=0,
            \end{aligned}\label{eq:GFMM:dynamBoundary}
        \end{align}
        ensuring balance of normal stresses and zero shear, where $\mathbf{u}_i$ denotes the velocity of the interface, $\mathbf{n}=\frac{\nabla\varphi}{\|\nabla\varphi\|}$ and $\mathbf{t}$ are the surface normal and tangential vectors, $g$ is the gravitational acceleration, $\boldsymbol{\tau}$ is the stress tensor, and $\kappa=\frac{\nabla\cdot(\nabla\varphi)}{\|\nabla\varphi\|}$ is the surface curvature.
        The authors note that integrating \eqref{eq:GFMM:phieq1} does not guarantee that the thickness of the interface region remains constant in space and time due to various errors caused by numerical diffusion and distortion by the flow field, and employ a correction developed by Sussman et al.\cite{sussmanAdaptiveLevelSet1999}
        \item The micro-scale cavities are resolved in the Lagrangian framework, where the size and trajectories of bubbles either formed from nuclei present in the flow at the start of the simulation, from nuclei created during the simulation due to nucleation at a solid boundary, or from the breakup of larger cavities are all tracked.
        The size of the bubble is modelled through its radius $R$, which is approximated by a modified RPE given by
        \begin{equation}
            \begin{split}
                \rho_l\left(R\Ddot{R}+\frac{3}{2}\dot{R}^2\right)&=p_v+p_{g0}\left(\frac{R_0}{R}\right)^{3k}-p_{enc}\\
                &\hspace{3ex}-\frac{2\sigma}{R}-\frac{4\mu_l R}{R}+\rho_l\frac{\|\mathbf{u}_s\|^2}{4},\label{eq:GFMM:RPE}
            \end{split}
        \end{equation}
        where $p_{enc}$ is the averaged of the liquid pressure over the surface of the bubble, $p_{g0}$ is the initial gas pressure inside the bubble, $R_0$ is the initial bubble radius, $k$ is the gas polytropic compression constant, and $\mathbf{u}_s=\mathbf{u}_{enc}-\mathbf{u}_b$ is the velocity slip, defined here as the difference between the liquid velocity averaged over the surface of the bubble $\mathbf{u}_{enc}$ and the bubble's translation velocity $\mathbf{u}_b$; the last term in \eqref{eq:GFMM:RPE} accounts for the effects of slip velocity between the motion of the bubble and the flow of the surrounding liquid, and was first derived by Hsiao et al\cite{hsiaoScalingEffectBubble2000}.
        Additionally, the usage of surface-averaged flow characteristics in \eqref{eq:GFMM:RPE} was introduced by Hsiao et al. \cite{hsiaoScalingEffectPrediction2003} in order to account for the prescence of a non-uniform pressure field in the immediate vicinity of the bubble.
        The bubble's translation velocity $\mathbf{u}_b$ is obtained from an equation for the bubble motion given by
        \begin{equation}
            \begin{split}
                \frac{d\mathbf{u}_b}{dt}&=\left(\frac{\rho_l}{\rho_b}\right)\left[\frac{3}{8R}C_D\|\mathbf{u}_s\|\mathbf{u}_s+\frac{1}{2}\left(\frac{d\mathbf{u}_{enc}}{dt}-\frac{d\mathbf{u}_b}{dt}\right)\right.\\
                &\hspace{-1ex}+\left.\frac{3\dot{R}}{2R}\mathbf{u}_s-\frac{\nabla p}{\rho_l}+\frac{\rho_b-\rho_l}{\rho_l}\mathbf{g}+\frac{3C_L\sqrt{\nu}(\mathbf{u}_s\times \Omega)}{4\pi R\sqrt{\|\Omega\|}}\right].
            \end{split}\label{eq:GFMM:bubble_motion}
        \end{equation}
        The terms on the right-hand side of the bubble motion equation \eqref{eq:GFMM:bubble_motion} represent various contributions to changes in the bubble's trajectory due to drag forces, added mass of the bubble, the pressure gradient of the surrounding liquid, gravitational forces, and lift forces as expressed via the vorticity vector $\Omega$.
        Additionally, the expressions for the first and last terms were first developed by Haberman and Morton\cite{habermanExperimentalInvestigationDrag1953} and Saffman\cite{saffmanLiftSmallSphere1965}, respectively.
        \item The model features two transition schemes, one for the micro-macro transition and one for the macro-micro transition.
        The micro-macro transition scheme transitions a micro-scale bubble to the macro-scale if its current radius both exceeds a threshold value and is greater than a specified multiple of the local grid size as specified by the criterion
        \begin{equation}
            R\geq \max(R_{thr},m_{thr}\Delta L),\label{eq:GFMM:microMacro}
        \end{equation}
        where $R_{thr}$ is the threshold radius, $\Delta L$ is the size of the local grid hosting the bubble, and $m_{thr}$ is a threshold grid factor.
        When a bubble is detected to satisfy \eqref{eq:GFMM:microMacro}, the bubble is removed from the simulation, and a new micro-scale cavity is introduced in the cell(s) occupied by the bubble.
        With this scheme, the model is capable of simulating both multiple bubbles coalescing into a large cavity and single isolated bubbles coalescing with a previously defined large cavity.
        The macro-micro transition scheme instead determines if the shape of the macro-scale cavity, identified here as the zero level set of $\varphi$, has developed sufficient instabilities to breakup into a cloud of bubbles based on empirical criteria described by Ma et al.\cite{maTwofluidModelingBubbly2011} and Hsiao et al\cite{hsiaoNumericalExperimentalStudy2013}.
        Once a cavity satisfies these criteria, it is replaced with a cloud of micro-scale bubbles of uniform size.
        Additionally, the fluid is initialized with no macro-scale cavities present in the flow and with micro-scale nuclei randomly distributed in the flow based on previous experimental measurements by Medwin \cite{medwinCountingBubblesAcoustically1977}, Billet\cite{billetCavitationNucleiMeasurements1985}, Franklin\cite{franklinNoteRadiusDistribution1992}, and Wu and Chahine\cite{wuDevelopmentAcousticInstrument2010}.
        Additional nuclei are added at each time step to the flow from both the inlet using the same distribution as the initial distribution of nuclei and from bubble entrainment occurring at the solid boundaries.
    \end{enumerate}
    
    \red{The Ghost-Fluid Multiscale Model is an in-house development employing the finite-volume method with implicit time integration. The different phases are modelled with the level set method, or the Lagrangian discrete bubble model.
    The model was validated against experimental data of shedding frequencies and cavity length for cavitating flow over a hydrofoil reported by Berntsen et al.\cite{berntsenNumericalModelingSheet2001} (e14), showing good agreement.
    Ma et al.\cite{maPhysicsBasedMultiscale2017} used this model in a study of bubbly flow within a waterjet propulsion nozzle as well as unsteady sheet cavitation on a hydrofoil.
    }

\subsection{The Density-based Convex Combination Model}
    Huang and Wang's\cite{huangModifiedDensityBased2011} model is a TEM with source terms derived using a convex combination of the source terms from two previous models, namely the source terms from the Vapor Nuclei-Adjusted Model\cite{zwartTwophaseFlowModel2004} by Zwart et al.\cite{zwartTwophaseFlowModel2004} given by
    \begin{align}
        \Dot{m}_{1}^+ &= C_{+}\frac{3\alpha_{nuc}(1-\alpha_v)\rho_v}{R}\sqrt{\frac{p_v-p}{\rho_l}},& p&<\hspace{-0.3ex}p_v,\\
        \Dot{m}_{1}^- &= C_-\frac{3\alpha_v\rho_v}{R}\sqrt{\frac{p-p_v}{\rho_l}}, & p&>p_v
    \end{align}
    and the source terms from the Interface Mass and Normal Momentum Model\cite{senocakInterfacialDynamicsbasedModelling2004} given by
    \begin{align}
        \Dot{m}_{2}^+ &= \frac{\rho_l(p_v-p)\alpha_l}{\rho_v(u_{v,n}-u_{i,n})^2(\rho_l-\rho_v)t_\infty},& p&<p_v,\\
        \Dot{m}_{2}^- &= \frac{(p-p_v)(1-\alpha_l)}{(u_{v,n}-u_{i,n})^2(\rho_l-\rho_v)t_\infty},& p&>p_v.
    \end{align}
    in an attempt to combine the advantages of both models.
    The authors introduce their convex combination by first defining a \say{blending function} $\chi$ given by
    \begin{equation}
        \chi\left(\frac{\rho_m}{\rho_l}\right)=\frac{1}{2}+\frac{\tanh\left[\frac{C_1(0.6\rho_m/\rho_l - C_2)}{0.2(1-2C_2)+C_2}\right]}{2\tanh(C_1)},
    \end{equation}
    then using the source terms of both models to define
    \begin{align}
        \begin{aligned}
        \Dot{m}^+ &= \chi\left(\frac{\rho_m}{\rho_l}\right)\Dot{m}_{1}^+ \\
        &\hspace{2ex}+ \left(1-\chi\left(\frac{\rho_m}{\rho_l}\right)\right)\Dot{m}_{2}^+,
        \end{aligned}\quad p<p_v, \\
        \begin{aligned}
        \Dot{m}^- &= \chi\left(\frac{\rho_m}{\rho_l}\right)\Dot{m}_{1}^- \\
        &\hspace{2ex}+ \left(1-\chi\left(\frac{\rho_m}{\rho_l}\right)\right)\Dot{m}_{2}^-,
        \end{aligned}\quad p>p_v.
    \end{align}
    The weights of this convex combination are defined using a function of the local mixture density, lending greater weight to the source terms from the Vapor Nuclei-Adjusted Model as the mixture density approaches the pure liquid density.
    Conversely, the source terms from the Interface Mass and Normal Momentum Model are favored as the mixture density approaches the vapor density.
    \red{Huang and Wang do not report the method used for discretizing the Navier-Stokes equations, but mention that a modified $k-\varepsilon$ model was used for turbulence modelling.
    The model is validated by simulation cloud cavitation over a hydrofoil, showing good agreement with experimental data previously reported by Wang et al.\cite{wangDynamicsAttachedTurbulent2001} (e15).}

\subsection{The Heat Balance Model}
    Shi et al.'s\cite{shiRayleighPlessetBasedTransport2014} model expresses the source terms using a revised simplification of the RPE which has been modified to include thermal effects.
    The authors choose the model
    \begin{align}
        \Dot{m}^+ &=  C_{+}\frac{3\alpha_v\rho_v}{R}\sqrt{\frac{p_v-p}{\rho_l}},& p&< p_v\label{eq:heatModel:1.1}\\
        \Dot{m}^- &= C_-\frac{3(1-\alpha_v)\rho_v}{R}\sqrt{\frac{p-p_v}{\rho_l}}, & p&> p_v\label{eq:heatModel:1.2}
    \end{align}
    with empirical constants $C_{+}$ and $C_{-}$ as a starting point, based on the previous work of Merkle et al.\cite{merkleComputationalModelingDynamics1998}
    The new development in the Heat Balance Model is an extension of the approximation of the bubble dynamics given by the simplified RPE
    \begin{equation}
        \frac{dR}{dt}=\sqrt{\frac{|p_v-p|}{\rho_l}}\label{eq:heatModel:2}
    \end{equation}
    used in the expressions \eqref{eq:heatModel:1.1} and \eqref{eq:heatModel:1.2} to the case of cryogenic cavitating flows. 
    To this end, the authors consider the heat flux $q$ of a transiently evolving bubble, which at a time $t$ is given by
    \begin{equation}
        q=\frac{K\Delta T}{\sqrt{at}}\label{eq:heatModel:3}
    \end{equation}
    along with the heat balance across the bubble's interface, expressed as
    \begin{equation}
        4q\pi R^2=\frac{4\pi\rho_v}{3}\frac{d}{dt}(R^3).\label{eq:heatModel:4}
    \end{equation}
    In the expressions \eqref{eq:heatModel:3} and \eqref{eq:heatModel:4}, $K=a\rho_lc_p$ is the thermal conductivity, $a$ is the thermal diffusivity, $c_p$ is the specific heat at constant pressure, $L$ is the latent heat, and $\Delta T=|T_c-T_\infty|$ is the temperature drop expressed via the local temperature $T_c$ and the freestream temperature $T_\infty$.
    By combining \eqref{eq:heatModel:3} and \eqref{eq:heatModel:4}, the authors obtain the relation
    \begin{equation}
        \frac{dR}{dt}=\frac{\rho_l c_p\sqrt{a}\Delta T}{\rho_vL\sqrt{t}},
    \end{equation}
    which is added to the original approximation \eqref{eq:heatModel:2} of the bubble dynamics, yielding the modified expressions of the source terms given by
    \begin{align}
        \begin{aligned}
        \Dot{m}^+ &=  C_{+}\frac{3\alpha_v\rho_v}{R}\left(\sqrt{\frac{\Tilde{p}_v-p}{\rho_l}}\right.\\
        &\hspace{1.25ex}+\left.\frac{\rho_l c_p\sqrt{a}\max\{T_\infty-T_c,0\}}{\rho_vL\sqrt{t}}\right),\quad 
        \end{aligned}p<\Tilde{p}_v,\label{eq:heatModel:4.1}\\
        \begin{aligned}
        \Dot{m}^- &= C_-\frac{3(1-\alpha_v)\rho_v}{R}\left(\sqrt{\frac{p-\Tilde{p}_v}{\rho_l}}\right.\\
        &\hspace{1.25ex}+\left.\frac{\rho_l c_p\sqrt{a}\max\{T_c-T_\infty,0\}}{\rho_vL\sqrt{t}}\right),\quad 
        \end{aligned}p>\Tilde{p}_v,\label{eq:heatModel:4.2}
    \end{align}
    where the vapor pressure $\Tilde{p}_v$ is obtained by modifying the local vapor pressure $p_v(T_l)$ at temperature $T_l$ in the same manner as that of the Full Cavitation Model\cite{singhalMathematicalBasisValidation2002} in order to account for the local fluctuations in pressure induced by turbulence, i.e.
    \begin{equation}
        \Tilde{p}_v = p_v(T_l)+\frac{0.39\rho_m k}{2}.
    \end{equation}
    \red{The authors do not mention which flow discretization method was used.
    The turbulence was modelled using a shear stress model by  Strelets\cite{streletsDetachedEddySimulation2001}.
    The model was validated by simulating cavitating flow over an axisymmetric ogive previously investigated experimentally by Sarósdy and Acosta\cite{sarosdyNoteObservationsCavitation1961} (e16), showing good agreement with their findings.}
    % \blue{
    % No references to later applications of this model towards simulation were found in the literature; the model was cited by Guo et al.\cite{guoAnticavitationPerformanceSplitterbladed2015} in their study of anti-cavitation performance.
    % }

\subsection{The Population Balance Model}
    Li and Carrica\cite{liPopulationBalanceCavitation2021} derive the source terms using the coupling given by \eqref{eq:cavmodel_vaporvolfrac_expr} and a simplified version of the RPE \eqref{eq:rayleighPlesset}.
    The unique feature of this model is the treatment of the vapor volume fraction $\alpha_v$, which is calculated as
    \begin{align}
        \alpha_v=\sum_{g=1}^G \frac{m_g}{\rho_g}N_g,
    \end{align}
    where the distribution of bubbles in the cavitating flow is assumed to be distributed among $G$ groups of bubbles, in which every bubble in a group $g$ is assumed to have a uniform mass $m_g$ and constant density $\rho_g$, with $N_g$ denoting the bubble number density of group $g$.
    The total interphase mass transfer rate is given by
    \begin{align}
        \begin{aligned}
        \Dot{m}&=\sqrt{\frac{32}{3}}\pi\rho_vR^2\frac{p_v-p}{\sqrt{\rho_l\max\{|p_v-p|,\epsilon\}}}\\
        &\hspace{3ex}\times(1-0.5\tanh(20(\alpha_v-0.7))),
        \end{aligned}
    \end{align}
    where $\epsilon>0$ is a small positive number introduced to prevent division by zero.
    In order to model the evolution of the bubble number densities of the various groups, the authors consider a Boltzmann-like transport equation
    \begin{equation}
        \frac{\partial f}{\partial t}+\nabla \cdot(f\mathbf{u}_l)+\frac{\partial (\Dot{m}f)}{\partial t}=\beta +\chi + S,
    \end{equation}
    where $f$ is the bubble size distribution function and $\beta$, $\chi$, and $S$ are source terms due to bubble breakup, coalescence, and entrainment, respectively; the authors here note that $f$, $\beta$, $\chi$, and $S$ are all functions of the mass $m$.
    Using a multigroup approach to discretize the transport equation into a series of $G$ transport equations for the bubble number densities $N_g$, the authors aim to obtain expressions for the corresponding source terms $\beta_g$, $\chi_g$, and $S_g$ by extending previously developed models for these source terms in the case of noncavitating flows to cavitating flows.
    The source term for coalescence $\chi=\chi^+-\chi^-$ is expressed using an extension of the model originally developed in Prince and Blanch\cite{princeBubbleCoalescenceBreakup1990}, in which the production and destruction of bubbles of mass $m$ are given by
    \begin{align}
        \begin{aligned}
        \chi^+(m) = \frac{1}{2}\int_{0}^m &T(m-m',m')C(m-m',m')\\
        &\times f(m-m')f(m')\,dm',
        \end{aligned}
    \end{align}
    \begin{align}
        \begin{aligned}
        \chi^-(m) =\int_{0}^\infty &T(m,m')C(m,m')\\
        &\times f(m)f(m')\,dm'
        \end{aligned}
    \end{align}
    where $T$ and $C$ are the bubble collision rate and the coalescence efficiency rate between bubbles of mass $m$ and $m'$, respectively, which are modelled as
    \begin{align}
        \begin{aligned}
        T(m,m')&=\frac{(R_m+R_{m'})^2}{1-\alpha_v}\Big(\pi\|\mathbf{u}_m-\mathbf{u}_{m'}\|\\
        &+1.33(R_m + R_{m'})\|\nabla \mathbf{u}_m\|\\
        &+\left.1.41 (R_m^{2/3}+R_{m'}^{2/3})^{1/2}+\varepsilon^{1/3}\right),
        \end{aligned}
    \end{align}
    \begin{align}
        \begin{aligned}
        C(m,m')&=\exp\left(-\sqrt{\frac{(R_mR_{m'})^3\rho_l}{(R_{m}+R_{m'})^3 8\sigma}}\right.\\
        &\hspace{2ex}\times\left.\frac{2\ln\left(\frac{h_0}{h_f}\right) (R_m+R_{m'})}{2(\varepsilon(R_m+R_{m'}))^{1/3}+ \|\mathbf{u}_r\|}\right).
        \end{aligned}
    \end{align}
    To develop a model for the source term due to bubble breakup, the authors assume that bubbles may only break up into two smaller bubbles, then consider two different effects that induce bubble breakup: breakups induced by turbulence and breakup due to bubble fission when a bubble collapses.
    In both cases, the production term $\beta^+$ and the destruction term $\beta^-$ are given by
    \begin{align}
        \beta^+(m) &= \int_m^\infty f(m')h_t(m,m')b(m')\,\mathrm{d}m', \\
        \beta^-(m) &= b(m)f(m),
    \end{align}
    where $b(m)$ is the bubble breakup rate for a bubble with mass $m$ and $h(m,m')$ is the daughter size distribution for bubbles of mass $m$ given a bubble with mass $m'$.
    To model bubble breakup induced by turbulence $\beta_{t}=\beta_{t}^+ -\beta_{t}^-$, the authors use the model introduced in Lehr et al.\cite{lehrBubbleSizeDistributionFlow2002}, in which the production and destruction terms are given by
    \begin{align}
        \beta_{t}^+(m) &= \int_m^\infty f(m')h_t(m,m')b_t(m')\,\mathrm{d}m', \\
        \beta_{t}^-(m) &= b_t(m)f(m),
    \end{align}
    where $b_t$ and $h_t$ represent the bubble breakup rate and the daughter size distribution given by
    \begin{align}
        b_t(m)&= \frac{(D_m^\ast)^{5/3}}{2\tau}\exp\left(-\frac{\sqrt{2}}{(D_m^\ast)^3}\right),\\
        h_t(m)&= \frac{1}{m\sqrt{\pi}}\frac{\exp\left(-\frac{9}{4}\left(\ln(2^{2/5}D_m^\ast)\right)^2\right)}{1+\mathrm{erf}\left(\frac{3}{2}\ln\left(2^{1/15}D_{m'}^\ast\right)\right)};
    \end{align}
    here, $D_m^\ast=2R(\rho_l/\sigma)^{3/5}\varepsilon^{2/5}$ is a dimensionless bubble diameter, $\tau=(\sigma/\rho_l)^{2/5}\varepsilon^{-3/5}$ is a time scale, and $\mathrm{erf}$ is the error function.
    The model for breakup induced by bubble fission $\beta_{bf}=\beta_{bf}^+ -\beta_{bf}^-$ is based on that of Brennen\cite{brennenFissionCollapsingCavitation2002}, which proposes a model for bubble fission under the assumption that the bubbles remain spherical during collapse.
    In this model, the daughter size distribution $h_{bf}$ and the bubble breakup rate $b_{bf}$ are given by
    \begin{align}
        h_{bf}(m,m')&= n'\delta(m-m'/n'),\\
        b_{bf}(m)&= \frac{1.1}{R}\left(\frac{p_{tot}-p_v}{\rho_l}\right)^{1/2}H(n'-2),
    \end{align}
    where $H$ is the Heaviside function, $\delta$ is the Dirac delta function, and $n'$ is an integer describing the number of daughter bubbles.
    The authors do not develop a new model for the source term due to entrainment and instead refer to the models presented in Castro et al.\cite{castroMechanisticModelBubble2016} and Li et al.\cite{liLargescaleSimulationShip2019}
    \red{
    Li and Carrica\cite{liPopulationBalanceCavitation2021} implement their model in the general purpose CFD code REX, using a variety of RANS and LES approaches for turbulence modelling along with a finite differences scheme for discretizing the governing equations and a level set approach for free surface modelling.
    Validation on twisted hydrofoil, showing good agreement with both previous experimental data by Foeth \cite{foethStructureThreedimensionalSheet2008} (e17) as well as simulations performed by Asnaghi et al\cite{asnaghiImprovementCavitationMass2017}.
    The model was applied by Li and Carrica\cite{liNumericalStudyCavitating2023} in a numerical study of the cavitating flow over a backward facing step, obtaining predictions of the bubble number densities that illustrate the locations where smaller and larger bubbles are concentrated as well as the important factors in the formation of the shedding cloud.}

\subsection{The Euler-Lagrangian Multiscale Model}
    The Euler-Lagrangian Multiscale Model developed by Ghahramani et al. \cite{ghahramaniNumericalSimulationAnalysis2021} is a state-of-the-art multiscale model, being the culmination of a series of previous iterations of cavitation models capable of accounting for a range of different effects on cavitation.
    As detailed in Ghahramani et al.\cite{ghahramaniRealizabilityImprovementsHybrid2018}, the model can be split into a combination of three schemes, the first two of which are concerned with developing appropriate source terms for the liquid volume fraction of the cavitating flow at the two distinct scales.
    The liquid volume fraction at the macro-scale and the micro-scale are treated separately and denoted by $\alpha$ and $\beta$, respectively.
    At each of the scales, the homogeneous mixture hypothesis \eqref{eq:mixtureRhoMu} expressed in terms of the liquid volume fraction is assumed to hold, i.e. the mixture density $\rho_m$ and mixture viscosity $\mu_m$ are given by
    \begin{align*}
        \rho_m&=\alpha\rho_l+(1-\alpha)\rho_v, & \mu_m&=\alpha\mu_l+(1-\alpha)\mu_v,\\
        \rho_m&=\beta\rho_l+(1-\beta)\rho_v, & \mu_m&=\beta\mu_l+(1-\beta)\mu_v.
    \end{align*}
    \begin{enumerate}
        \item At the macro-scale, the flow characteristics such as velocity and pressure along with turbulence are modelled by using large-eddy simulation, and the source term in the transport equation \eqref{eq:vof} for $\alpha$ is modelled using a modified version of the Bubble Density-Liquid Volume Coupling Model\cite{schnerrPhysicalNumericalModeling2001} based on previous results from both Schenke and Terwisga\cite{schenkeSimulatingCompressibilityCavitating2017} as well as Ghahramani et al.\cite{ghahramaniComparativeStudyNumerical2019}.
        
        \item At the micro-scale, the vapor phase is tracked by tracking the size and location of parcels of (spherical) vapor bubbles, where each parcel consists of bubbles of similar radius, and the liquid volume fraction $\beta$ is instead obtained by estimating the corresponding vapor volume fraction $1-\beta$ on a cell-by-cell basis.
        This is done by determining the number of bubbles $n_i$ of a parcel of bubbles with the same radius $R_i$ that occupy a given cell with index $j$, estimating the volume fraction of cell $j$ occupied by parcel $i$, and finally summing up the volume contributions from each of the parcels of bubbles $i=1,\dotsc,N_{b,j}$ occupying cell $j$.
        This approach to estimating the liquid volume fraction necessitates tracking the motion of each parcel of bubbles as expressed by their position $\mathbf{x}_b$ and velocity $\mathbf{u}_b$ as well as the sizes of the bubbles as expressed by their radii $R_i$.
        Given the mass $m_b$ of the bubbles, the equations governing the bubbles' motion is expressed in the Lagrangian framework as
        \begin{align}
            \frac{d\mathbf{x}_b}{dt}&=\mathbf{u}_b,\label{eq:multiscale_3.1}\\
            m_b\frac{d\mathbf{u}_b}{dt}&=\mathbf{F}_d+\mathbf{F}_l+\mathbf{F}_a+\mathbf{F}_p+\mathbf{F}_b+\mathbf{F}_g,\label{eq:multiscale_3.2}
        \end{align}
        where the various terms on the rhs. of \eqref{eq:multiscale_3.2} represent force components due to sphere drag force, lift force, added mass, pressure gradient force, buoyancy force, and gravity, respectively.
        The drag forces are expressed via a drag coefficient $C_D$ first derived by Amsden et al.\cite{amsdenKIVAIIComputerProgram1989}, and the lift forces are similarly expressed using a lift coefficient $C_l$ first derived by Mei\cite{meiApproximateExpressionShear1992}.
        The evolution of the bubble radius is governed by a modified RPE including the effects of surface tension and non-condensable gas given by
        \begin{align}
            \begin{aligned}
                \frac{1}{2}R\Ddot{R}+\frac{17}{32}\dot{R}^2&=\frac{p_v-p_{2R}}{\rho_l}+\frac{p_{g0}}{\rho_l}\left(\frac{R_0}{R}\right)^{3k}\\
                &\hspace{3ex}-\frac{4\mu_l\dot{R}}{\rho_lR}-\frac{2\sigma}{\rho_lR},
            \end{aligned}\label{eq:multiscale_7}
        \end{align}
        where $p_{2R}$ denotes the surface-average pressure of the mixture over a concentric sphere of radius $2R$ and replaces the freestream pressure $p_\infty$ in the RPE.
        This modification was derived in Ghahramani et al.\cite{ghahramaniComparativeStudyNumerical2019} under the assumption that the surface-averaged pressure $p_{2R}$ gives a better representation of the behavior of the pressure field in the immediate vicinity of the bubble.
    
        \item The transition scheme of the model determines if a cavity tracked in the Eulerian macro-scale scheme should transition to the Lagrangian micro-scale scheme or a cavity tracked in the Lagrangian micro-scale scheme should transition to the Eulerian macro-scale scheme by considering the number of computational cells used to represent the cavity; note here that a Lagrangian cavity refers to a cloud of any amount of micro-scale bubbles.
        Two threshold values on the number of cells $N_{EL}$ and $N_{LE}$ are used to form the criteria for the transition schemes, chosen such that $N_{LE}>N_{EL}$.
        If the number of cells used to represent a macro-scale cavity is less than $N_{EL}$, it is transitioned to a micro-scale cavity; otherwise it is kept in the Eulerian framework.
        In a similar manner, if the number of cells used to represent a (cloud of) micro-scale bubble(s) exceeds $N_{LE}$, the (cloud of) bubble(s) is transitioned to a macro-scale cavity in the Eulerian framework; otherwise it is kept in the Lagrangian framework.
        In both cases, the micro-/macro-scale cavities that satisfy the criteria are replaced by corresponding macro-/micro-scale cavities that occupy the same amount of volume in the fluid.
    
        Furthermore, collisions between two distinct cavities as well as turbulence-induced breakage of cavities are also modelled, along with corrections to the mixture properties and mass transfer rates due to cavities transferring from one scheme to the other.
        Collisions between micro-scale bubbles and macro-scale cavities are modelled by absorbing the micro-scale bubble into the macro-scale cavity, whilst collisions between micro-scale cavities are modelled in two steps using a model introduced by Breuer and Alletto\cite{breuerEfficientSimulationParticleladen2012} and further extended by Vallier \cite{vallierSimulationsCavitationLarge2013} to first detect incidence of collisions based on the current trajectories of the cavities, then determining whether the colliding cavities remain in contact for long enough to coalesce based on characteristic time scales derived by Kamp et al.\cite{kampBubbleCoalescenceTurbulent2001} or bounce back from each other.
        Finally, the turbulence-induced breakage of cavities is modelled using a criterion introduced by Lau et al.\cite{lauNumericalStudyBubble2014} and Hoppe and Breuer\cite{hoppeDeterministicBreakupModel2020}.
        This criterion declares a bubble of diameter $d_p$ undergoes breakage if its Weber number
        \begin{equation}
            \text{We}=\frac{\rho_l\overline{(u_i'u_i')_{d_p}}d_p}{\sigma},
        \end{equation}
        exceeds the critical value $\text{We}=15.12$, where $\overline{(u_i'u_i')_{d_p}}$ is the mean square velocity difference over a distance equal to the diameter of the bubble; this choice of critical value corresponds to assuming that only binary breakups into two equally-sized daughter bubbles of diameter $d_s$ occur.
        According to Hoppe and Breuer\cite{hoppeDeterministicBreakupModel2020}, this implies that the ratio $\frac{d_p}{d_s}$ is equal to 1.26, and this relation can be used to determine the corresponding bubble radius $R_s$ of the daughter bubbles.
    \end{enumerate}
    \red{
    Ghahramani et al. \cite{ghahramaniNumericalSimulationAnalysis2021} implement their model by combining their micro-scale Lagrangian model and transition scheme with the interPhaseChangeFoam solver implemented in the open software package OpenFOAM.
    Validation against exp. data of periodic cavitating flow over a bluff body reported by Ghahramani et al. \cite{ghahramaniExperimentalNumericalStudy2020} (e18), showing good agreement.
    Brandner et al. \cite{brandnerNucleationEffectsCavitation2022} performed experimental investigations of nucleation effects on cavitation effects about a sphere with the stated aim of providing a high-fidelity dataset for further improvements of the micro-scale modelling employed in this model.
    }

\subsection{The Stochastic Field Model}
    Due to the close connection between cavitation and turbulence demonstrated by several authors \cite{ohernExperimentalInvestigationTurbulent1990,brandnerExperimentalInvestigationCloud2010,huangLargeEddySimulation2014}, it is natural to seek a stochastic description of cavitation, as turbulence is by definition a stochastic phenomenon.
    These considerations have lead to the development of stochastic cavitation models, which aim to construct models for quantities such as the volume fraction using probabilistic models.
    This approach involves formulating an appropriate stochastic process that describes the process of cavitation as well as a stochastic partial differential equation that governs the evolution of said process, a similar procedure to the approach by deterministic cavitation models such as TEMs.
    Once this model has been established, the cavitation process can be simulated by using tools developed for stochastic partial differential equations to obtain a realization of the desired stochastic process.
    The most prominent stochastic model developed thus far is the Stochastic Field Model due to Dumond et al. \cite{dumondStochasticfieldCavitationModel2013}, which models the probability density function $f_Y$ of the vapor mass fraction, denoted here by $Y$, by applying the stochastic field method previously developed by Valiño \cite{valinoFieldMonteCarlo1998}.
    Within this method, the pdf $f_Y$ is approximated as a sum of stochastic fields $Y^k$ as follows:
    \begin{equation}
        f_Y(y;\mathbf{x},t)\approx\frac{1}{N}\sum_{k=1}^N \delta(y-Y^k(\mathbf{x},t))\label{eq:stoc_1}
    \end{equation}
    These stochastic fields $Y^k$ represent possible realizations of the true vapor mass fraction $Y$, and in the limit as $N\rightarrow\infty$, the approximation \eqref{eq:stoc_1} converges to the true pdf $f_Y$.
    For practical applications, the authors recommend the value $N=8$ for a good compromise between stability and efficiency based on similar models developed for simulating combustion.
    Each field $Y^k$ is determined in practice by solving its associated stochastic partial differential equation, which the authors develop in the Itô calculus using the methods of Gardiner\cite{gardinerHandbookStochasticMethods1983} as
    \begin{equation}
        \begin{split}
            dY^k&=-u_i\frac{\partial Y^k}{\partial x_i}dt+S(Y^k)dt+\frac{\partial}{\partial x_i}\left(D_Y'\frac{\partial Y^k}{\partial x_i}\right)dt\\
            &\hspace{3ex}+\sqrt{2D_Y'}\frac{\partial Y^k}{\partial x_i}dW_i^k-\frac{Y^k-\langle Y \rangle}{2\tau_Y},
        \end{split}
    \end{equation}
    where $\langle Y \rangle=\frac{1}{N}\sum_{k=1}^N Y^k$ is the average of the stochastic fields, $D_Y'$ and $\tau_Y$ are respectively the diffusivity coefficient and the turbulent relaxation time obtained from a turbulence model, the $W_i^k$ are Wiener processes that are independent for each $i$ and constant in space, meaning that their time derivatives $dW_i^k$ are independent, normally distributed variables with zero mean and unity variance and can thus be obtained from a random number generator, and $S(Y^k)$ is a source term.
    This source term $S(Y^k)$ is split into two terms $S(Y^k)=S_+(Y^k)-S_-(Y^k)$, each of which is modelled using the RPE \eqref{eq:rayleighPlesset} to express the evolution of the bubble radius
    $R^k=\left(\frac{3\rho Y^k}{4\pi\rho_g n}\right)^{1/3}$
    of the $n$ cavities per unit volume, where $n$ is defined using the vapor volume fraction $\alpha_v=\frac{\rho\langle Y \rangle}{\rho_g}$ and a specified initial number of nuclei $n_0$ as
    \begin{equation}
        n(\alpha_v)= \frac{n_0+1}{2}+\frac{n_0-1}{2}\tanh\left(5(\alpha-0.4)\right).
    \end{equation}
    Using the bubble radius $R^k$, the authors derive the following expressions for the source terms $S_+(Y^k)$ and $S_-(Y^k)$:
    \begin{align*}
        S_+(Y^k)&= (36n\rho_g\pi)^{1/3}(\rho Y^k)^{2/3}\\
        &\hspace{2ex}\times\sqrt{\frac{2}{3\rho_l}\max\left(p_v(T)-p-\frac{4\sigma}{3R^k},0\right)},\\
        S_-(Y^k)&=\begin{cases}
        S_-^1(Y^k) & Y^k\leq Y_{eq}\\
        S_-^2(Y^k) & Y^k> Y_{eq}
        \end{cases},
    \end{align*}
    where
    \begin{align*}
        S_-^1(Y^k)&= \frac{\rho}{\tau_{nuc}}(Y_{eq}-Y^k),\\
        S_-^2(Y^k)&= -C_{cond}(36n\rho_g\pi)^{1/3}(\rho Y^k)^{2/3}\\
        &\hspace{2ex}\times\sqrt{\frac{2}{3\rho_l}\max\left(p_v(T)-p,0\right)}
    \end{align*}
    and $\tau_{nuc}$ and $C_{cond}$ are both modelling constants that ensure vaporous cavities do not become smaller than initial nuclei after collapse and account for inertial effects on condensation, respectively.
    Furthermore, the pressure of the mixture is obtained using the EOS derived by Okuda and Ikohagi\cite{okudaNumericalSimulationCollapsing1996}
    \begin{equation}
        \rho_m=\frac{p(p+p_c)}{K(1-f_v)p(T+T_c)+Rf_v(p+p_c)T},
    \end{equation}
    where $p_c$, $T_c$, $K$, and $R$ are the pressure, temperature, liquid, and gas constant of the fluid, respectively.
    The authors validate their model on test cases concerning Venturi nozzles and fluidic diodes, obtaining results that demonstrate their model's capability of replicating the cavitating flow in these cases, both quantitatively in terms of the predicted velocity profiles and vapor mass fractions that agree with experimental data obtained from Barre et al.\cite{barreExperimentsModelingCavitating2009} and Stutz and Reboud\cite{stutzMeasurementsUnsteadyCavitation2000} for the Venturi nozzle case, but also demonstrate detailed qualitative behaviour for both forwards and backwards flow in the fluidic diode case.
    \red{
    Dumond et al. \cite{dumondStochasticfieldCavitationModel2013} implemented their model in the large eddy simulation framework SPARC, using an explicit fourth order Runge-Kutta scheme for temporal discretization and the SWITCH central difference scheme with artificial dissipation for spatial discretization.
    Chen and Oevermann \cite{chenEulerianStochasticField2018} applied the Stochastic Field Model to simulate cavitating flow through a throttle inside a diesel injector, obtaining results that agree with previous numerical results by Altimira and Fuchs\cite{altimiraNumericalInvestigationThrottle2015}, who performed their simulations using the Bubble Density-Liquid Volume Coupling Model\cite{schnerrPhysicalNumericalModeling2001} and obtained results in agreement with experimental results previously reported by Winklhofer et al.\cite{winklhoferComprehensiveHydraulicFlow2001} (e19).
    }

\section{Conclusions}{
%     This review article presents a summary of the main classes of models developed for modelling cavitation, a multiphase phenomenon in which a fluid locally experiences phase change due to drops in pressure.
% Several examples of cavitation models are presented, which are then classified according to the approach used by a given model to model cavitation.
% For each of the models presented, the various assumptions and simplification made by the authors of the model is discussed, and applications of the model to simulating various aspects of cavitating flow are also presented.

% The main class of models identified in this review is that of the homogeneous cavitation models, in which cavitating flows are modelled as a homogeneous mixture of the two pure phases of the fluid. These models are further classified according to the approach used for employing this homogeneous mixture hypothesis.
% Other classes of cavitation models based by other approaches such as tracking the interface of the cavities or employing stochastic models are also identified.

% Using the preceding discussion of the various cavitation models presented, the review concludes with an outlook towards future improvements in the modelling of cavitation.   
\red{
    This review of cavitation modelling was developed as a supplement to the previous reviews in the literature, focusing on the physical implications of the approaches and assumptions employed by the authors of the models.
    Through analysis of the various proposed cavitation models, five commonly used approaches to modelling cavitation were identified, namely bubble dynamics as expressed via the Rayleigh-Plesset equation (RPE), direct simulation of mass transfer via transport equations (TEM), identifying cavities through an equation of state, relating fluid characteristics to thermodynamic variables (EOS), directly tracking the interfaces between the vapor phase and the liquid phase (ITM), and simulating multiple scales of cavities at once with separate schemes (MUL). 
    Figure \ref{fig:cav_model_venn} indicates how the different models are composed of the different modelling approaches. 
    Starting clockwise from the upper-left of Figure \ref{fig:cav_model_venn}, the highlighted models have been labelled from 1 to 20 starting with models only belonging to a single category, followed by labelling models belonging to two categories clockwise from the upper-left, and so forth.
    In this representation, simple, fundamental models lie on the outside, while more complex models, which involve several modelling approaches, are found towards the centre.
}

    \begin{figure}
    \centering
    \includegraphics[width=0.475\textwidth]{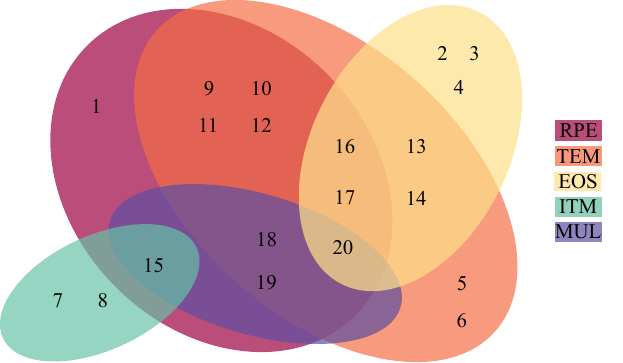}
    \caption{The cavitation models highlighted in this review, categorized by the approach employed by the model using the categories described in section \ref{subsec:approach}.}
    \label{fig:cav_model_venn}
    \end{figure}

\red{       
    Kubota et al.'s\cite{kubotaNewModellingCavitating1992} Bubble Cluster Model is frequently referred to as inspiration for RPE models. 
    The Bubble Cluster Model exclusively models the cavitation dynamics by developing an equation for the motion of a cluster of bubbles.
    While Kubota et al.'s approach is appealing from a physical perspective, the model development has obviously favored to embed all phase transitions and bubble dynamics in the source term of a transport equation of the vapor volume fraction.
    These resultant TEM-RPE models pose the largest group of models, see Figure \ref{fig:cav_model_venn}.    
    Based on the preceding analysis and categorization, some general conclusions can be drawn related to the modelling approach: 
    }

\red{
    \begin{itemize}
        \item The current standard for cavitation modelling is set by the combination of TEM and RPE approaches. The most widely applied models are proposed by authors such as Kunz et al.\cite{kunzPreconditionedNavierStokes2000}, Schnerr and Sauer\cite{schnerrPhysicalNumericalModeling2001}, Singhal et al.\cite{singhalMathematicalBasisValidation2002}, and Zwart et al.\cite{zwartTwophaseFlowModel2004}, which all date back to around the turn of the millennium.
        \item These standard models are all formulated within the volume-of-fluid framework, allowing them to be readily implemented into most CFD workflows, while being computationally efficient.
        As such, they have been adopted both by a large proportion of the scientific community studying cavitation, but also within both open-source software packages such as OpenFOAM and commercial software such as Ansys Fluent or STAR-CCM+.
        Despite almost all of these models being 20 years old, they continue to be the standard for most studies involving cavitation modelling.
        \item  EOS models incorporate thermodynamic phenomena in cavitating flow modelling. 
        In particular, the assumption of iso-thermal flow can be liberated.
        EOS models are rather applied to high resolution cases, indicating that they might be computationally more demanding.
        \item ITM models, on the other hand, allow for high-resolution simulations of the cavity interface dynamics.
        These models are utilized for studying two-dimensional problems only; a generalization to three dimensions seems to be problematic. 
        The application of ITMs is therefore limited to very specific problems only.        
        \item The MUL models developed more recently, e.g. the Euler-Lagrangian Multiscale Model by Ghahramani et al.\cite{ghahramaniNumericalSimulationAnalysis2021}, come closer to the realization of tracking the growth and motion of bubbles of similar size, in the fashion of Kubota's original Bubble Cluster Model. 
        For resolving larger cavities or dense bubble clouds, they still resort to the TEM approach.
    \end{itemize}
}

\red{
    Furthermore, four effects commonly accounted for in physical modelling of cavitation and implemented through modifications to the expressions for e.g. mass transfer rates were also identified as another way of categorizing cavitation models; these effects include empirical adjustment of the model to obtain a better fit (EMP), including turbulent effects on the pressure fluctuations or the deformation of cavities (TUR), accounting for changes in the population balance of cavities due to e.g. breakup of cavities or coalescence (POP), and allowing for the possible presence of non-condensable gas in the fluid (NCG).  
    Figure \ref{fig:cav_modelEffects_venn} illustrates the effects accounted for in the construction of the highlighted models.
    Based on the previous analysis, supported by Figure \ref{fig:cav_modelEffects_venn}, the following conclusions can be drawn on the model effects:
}
    % All models include some parameter adjusted according to the characteristics of the flow to be simulated, and the majority also account for the effects of turbulence. 
    
    % However, some models have become the standard cavitation methods in the commercial finite volume packages Ansys Fluent (M9-M11) and STAR-CCM+ (M9), and the open-source alternative OpenFOAM (M5,M9). 
    % Interestingly, with the exception of model 5, all these models simply combine an RPE-derived source term in a TEM framework, see Figure \ref{fig:cav_model_venn}.
    % Thus, the major CFD packages seem to rely on traditional, about 20 year old models for cavitation, while more modern approaches, such as multi-scale (MUL) and stochastic models are limited to scientific codes.

    \begin{figure}
    \centering
    \includegraphics[width=0.475\textwidth]{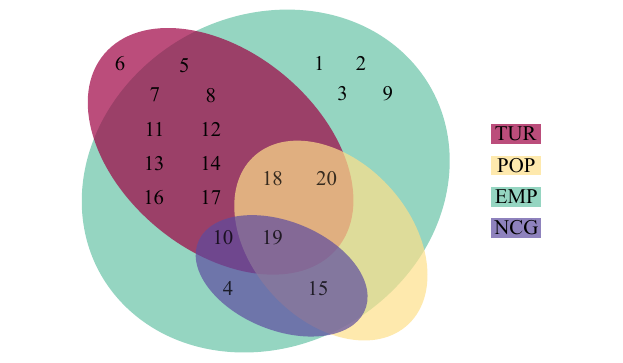}
    \caption{The cavitation models highlighted in this review, categorized by the effects accounted for by the model using the categories described in section \ref{subsec:effects}.}
    \label{fig:cav_modelEffects_venn}
    \end{figure}
    
\red{
    \begin{itemize}
        \item The majority of the models account for turbulence in some manner. 
        Many different turbulence models have been employed in the selected models: Reynolds-averaged approaches like the algebraic Baldwin-Lomax model\cite{baldwinThinlayerApproximationAlgebraic1978}, the one-equation Spalart-Allmaras model\cite{spalartOneEquationTurbulenceModel1992}, the two-equation $k-\varepsilon$ models, and shear stress models are prominently featured as the turbulence models of choice, but alternatives like large eddy simulation have also been employed successfully.
        No highlighted model has employed a turbulence model developed for mutliphase flows, opting for the established turbulence models which were developed for single-phase flows.
        \item All of the highlighted models feature adjustable parameters which reflect hypotheses placed on the fluid in question, e.g. the use of a prescribed uniform bubble number density or a characteristic time scale.
        The vast majority also feature empirical parameters used for calibrating the model to a specific instance of cavitating flow, much like turbulence models such as the $k-\varepsilon$ model.
        Only the Interface Mass and Normal Momentum Model proposed by Senocak and Shyy\cite{senocakInterfacialDynamicsbasedModelling2004} features no such empirical parameters, reflected in Figure \ref{fig:cav_modelEffects_venn} by its placement outside the category EMP.
        \item The models contained in the category MUL featured in Figure \ref{fig:cav_model_venn} are the same models contained in the category POP featured in Figure \ref{fig:cav_modelEffects_venn}, highlighting the fact that accounting for the effects on cavitation occurring at both the largest and the smallest scales necessitates accounting for the bubble population in the modelling approach.
        Furthermore, the majority of the models contained in this category have been developed in recent years, after the importance of accounting for the micro-scale effects was discovered.
        \item The category NCG contains less than a quarter of the highlighted models, indicating that most authors did not consider the effect of non-condensable gas relevant for their purposes and chose to neglect it.
    \end{itemize}
    }

}\label{sec:conclusion}

\section{Outlook}{

    \red{
    As noted in the introduction, there is currently no universal cavitation model, and the existence of such a model is predicated on obtaining a deeper understanding of other phenomena in fluid dynamics, including turbulence.
    To this end, the categorization presented in this article will assist in future investigations concerning the nature of cavitation as a phenomenon along with identifying which approach may be most appropriate for constructing a cavitation model given a system of interest.}
    \red{
    The categorization proposed in this work is neither exhaustive nor complete; future extensions and refinements of the categorization presented in this article are also expected in tandem with new developments in the various fields where cavitation occurs, both in terms of categorizing the modelling approach and the included effects.
    Additional aspects of cavitation modelling not considered here such as models for cavitation erosion will be the subject for future work and may entail an extension of the proposed categorization.
    }
    
    \red{
    Furthermore, the present discussion on cavitation models has been restricted to the physical modelling involved in their formulation, and attention to the software implementation and performance of cavitation models kept to a minimum (necessarily).
    A future comparative study on the performance of a selection of cavitation models when applied to test cases such as those highlighted in Table \ref{tab:expCases} will enrich this discussion by showcasing the adaptability of the models and quantifying accuracy and computational performance.
    }
    
    \red{
    It has been shown that many cavitation models are tightly coupled to concurrent turbulence models.
    However, the question of how the empirical tuning parameters relate to turbulence in the system remains open.
    As such, a sensitivity analysis of the empirical parameters against different turbulent regimes would promote the understanding of this correlation.
    Furthermore, since both cavitation and turbulence affect the macroscopic flow variables, it remains an open question if these two effects must not necessarily be modelled together.
    The findings of this sensitivity study may motivate the development of a new model focused on accounting for the inter-dependency of turbulence as thoroughly as possible.
    }
    % focus on the software implementation of selected cavitation models and investigate both their performance in a variety of test cases as well as characteristics such as grid independence.
    % \red{Analyses of empirical parameters, and their inter-dependency between turbulence modelling and cavitation. Sensitivity analysis using e.g. ML methods. Motivation for new turbulence-cavitation model combining effects on macroscopic variables.}
    
    \red{
    The formation and distribution of vapor nuclei in cavitating flows is handled via simplifying assumptions such as a uniform bubble number density per unit volume in many cavitation models; however, the population balance approach proposed by Li and Carrica\cite{liPopulationBalanceCavitation2021}, who noted the lack of cavitation experiments reporting bubble size data in their concluding remarks indicate that the subject is poorly understood.
    Other authors have performed experimental studies with the explicit aim of remedying this lack of data, e.g. the work of  Brandner et al. \cite{brandnerNucleationEffectsCavitation2022} detailing the influence of nucleation on cavitation about a sphere.
    An investigation into the stochastic properties of the vapor nuclei distributions typical of cavitating flows based on such data may shed further light on this subject and motivate further development of multiscale cavitation models.}

    \red{The phenomenon of cavitation in non-Newtonian fluids has been documented by Brujan and Williams\cite{brujanCavitationPhenomenaNonNewtonian2006,brujanCavitationBubbleDynamics2009,brujanCavitationNonNewtonianFluids2011}, but little effort towards developing cavitation models specifically for non-Newtonian fluids appears to have been made.
    A future study into modelling cavitation in non-Newtonian fluids, potentially using the framework of fractional Navier-Stokes equations as proposed by Zhou and Peng\cite{zhouTimefractionalNavierStokes2017}, would promote greater understanding of the nature of cavitation as a phenomenon.}
   
    \red{The increasing popularity of artificial intelligence and machine learning as a method for solving non-linear problems has led to multiple authors investigating the possibility of applying these tools to the problem of modelling cavitation.
    Some authors have developed methods that build upon previously established cavitation models, such as Sikirica et al. \cite{sikiricaCavitationModelCalibration2020}, who proposed a workflow for calibrating empirical constants of cavitation models for optimal performance using a random forest method, or Ouyang et al. \cite{ouyangReconstructionHydrofoilCavitation2023}, who applied a chain of physics-informed neural networks implementing both the Navier-Stokes equations governing fluid flow as well as a cavitation model of choice to provide better predictions of the changes in pressure that induce changes in the (liquid) volume fraction.
    Other approaches to both modelling cavitation and predicting the onset and intensity of cavitation using artificial intelligence have also been investigated; Xu et al. \cite{xuRANSSimulationUnsteady2021} proposed a method for enhancing the performance of Reynolds-averaged Navier-Stokes methods applied to modelling cavitating flows by improving the predicted values of the turbulent eddy viscosities using a random forest method, whilst Sha et al. \cite{shaMultitaskLearningCavitation2022} proposed a multi-task framework for detecting and classifying the intensity of cavitation via the acoustic signals emitted by the cavitating system using a bespoke neural network.
    Despite the advances made within the studies of both machine learning and cavitation in recent years, the application of machine learning to cavitation modelling has mostly been limited to optimizing previous approaches to modelling cavitation; a completely novel approach to modelling cavitation enabled by machine learning has not yet been discovered.
    Future work will investigate both existing applications of machine learning to cavitation modelling in greater detail as well as the possibility of devising a new approach to cavitation modelling enabled by machine learning.
    }

}\label{sec:outlook}

\section*{Data Availability Statement}
Data sharing is not applicable to this article as no new data were created or analyzed in this study.

\section*{Author Declarations}
The authors have no conflicts to disclose.

%-------------------------------------------------------------------
%-------------------------------------------------------------------

% If in two-column mode, this environment will change to single-column format so that long equations can be displayed. 
%\begin{widetext}
%$$\mbox{put long equation here}$$
%\end{widetext} 

% Figures should be put into the text as floats. 
% Use the graphics or graphicx packages (distributed with LaTeX2e).
% Put the label that you will use with \ref{} command in the braces of the \label{} command.
%
% \begin{figure}
% \end{figure}

% \begin{table}
% \end{tabular}
% \end{table}

%\begin{acknowledgments}
% Put your acknowledgments here.
%\end{acknowledgments}

\section*{References}
\bibliography{cavitation}
\end{document}